\documentclass[%
 reprint,
superscriptaddress,
nofootinbib,
 amsmath,amssymb,
 aps,
]{revtex4-2}

\usepackage{graphicx}%
\usepackage{dcolumn}%
\usepackage{bm}%
\usepackage{subcaption}
\usepackage{siunitx}

\usepackage{physics}
\begin{document}

\preprint{APS/123-QED}

\title{Iterative configuration of programmable unitary converter\\ based on few layer redundant multi-plane light conversion}

\author{Yoshitaka Taguchi}
\email{ytaguchi@ginjo.t.u-tokyo.ac.jp}
\affiliation{Department of Electrical Engineering and Information Systems, The University of Tokyo, 7-3-1 Hongo, Bunkyo-ku, Tokyo 113-8656 Japan}

\author{Yunzhuo Wang}
\affiliation{Preferred Networks Inc. Otemachi Bldg., 1-6-1 Otemachi, Chiyoda-ku, Tokyo 100-0004 Japan}

\author{Ryota Tanomura}
\affiliation{Department of Electrical Engineering and Information Systems, The University of Tokyo, 7-3-1 Hongo, Bunkyo-ku, Tokyo 113-8656 Japan}

\author{Takuo Tanemura}
\affiliation{Department of Electrical Engineering and Information Systems, The University of Tokyo, 7-3-1 Hongo, Bunkyo-ku, Tokyo 113-8656 Japan}

\author{Yasuyuki Ozeki}
\affiliation{Department of Electrical Engineering and Information Systems, The University of Tokyo, 7-3-1 Hongo, Bunkyo-ku, Tokyo 113-8656 Japan}

\date{\today}%

\begin{abstract}
Programmable unitary photonic devices are emerging as promising tools to implement unitary transformation for quantum information processing, machine learning, and optical communication. These devices typically use a rectangular mesh of Mach-Zehnder interferometers (MZIs), which has a clear mathematical structure and can be configured deterministically. However, this mesh architecture is sensitive to fabrication errors, and the correction techniques are still under investigation. In contrast, the multi-plane light conversion (MPLC) architecture is more robust against fabrication errors, but a deterministic method for configuring the converter has not yet been developed due to its complex mathematical structure. In this work, we propose a fast iterative configuration method for MPLC, following the mathematical review of the matrix distance and proposal of a new norm. We show through numerical simulations that adding a few redundant layers significantly improves the convergence of the MPLC architecture, making it a practical and attractive option. We also consider the effects of finite resolution and crosstalk in phase shifters in our simulations. In addition, we propose a phase-insensitive distance suited for applications using only intensity detections. Our method demonstrates orders of magnitude better accuracy and a 20-fold speed-up compared to previous approaches.
\end{abstract}

\maketitle

\section{Introduction}
\label{sec:intro}
Programmable unitary transformations implemented on integrated photonic platforms are becoming a powerful tool for a variety of applications, including quantum photonics \cite{Carolan2015,Wang2020,Elshaari2020,Carolan2020,Pelucchi2022,Chi2022,Madsen2022}, machine learning \cite{Shen2017,Prabhu2020,Zhang2021,Pai2022,Ashtiani2022,Ohno2022,Saumil2022}, and optical communication \cite{Fontaine2012,Annoni2017,Melati2017,Choutagunta2020,Tanomura2022ECOC}.
Accurate realization of a given unitary transformation is critical, as the fidelity of computational results and the error of optical communication can be significantly affected by the precision of the realized transformation.
A common approach to synthesizing unitary transformations is to use a mesh of Mach-Zehnder interferometers (MZIs) known as the Clements architecture \cite{Clements2016}, which consists of phase shifters and beam splitters (BSs). This architecture is attractive because its mathematical structure is decomposable, allowing the required phase shift in each MZI to be explicitly determined from the given unitary transformation.
However, physical implementation artifacts such as deviation in the splitting ratio of BSs can result in errors in the synthesized transformation. These errors can become significant as the number of optical modes increases \cite{Roel2017}. Several design proposals have been made in an effort to reduce or eliminate this error, with the goal of achieving a precise, customizable, and fabrication-error-tolerant unitary transformation that can be applied to scalable and reliable applications.

To address the challenge of implementation artifacts in the Clements architecture, several approaches have been proposed. One approach is local error correction, which involves fixing each MZI and can be applied to any MZI-based architecture, but requires prior knowledge of passive and active components \cite{Bandyopadhyay2021}.
Another approach is the measurement of components with on-chip power monitors, which allows for the calibration of each MZI but also increases the size of the chip and the complexity of wiring \cite{Miller2013,Miller2017}.
Self-configuration and 3-MZI approaches utilize an additional BS to achieve partially perfect linear operation and employ a feedback loop to adjust each phase shift using only output signals \cite{Hamerly2022Nat,Hamerly2022PRA}. While this method allows for infinite scalability, it also increases the size of the circuit and may have issues with stability \cite{Hamerly2022PRA2}. It's worth noting that these approaches primarily consider the artifacts of passive BSs in the circuit, and do not sufficiently consider the artifacts of phase shifters, such as crosstalk.

Another architecture employs a series connection of phase shifter arrays and unitary transformations to achieve a highly robust universal synthesis of unitary matrices that is resistant to fabrication errors. This architecture, also known as the multi-plane light conversion (MPLC) architecture \cite{Morizur2010,Labroille2014,Tang2017,Tang2018}, is particularly robust because each unitary transformation can be selected from a wide range of possible unitaries \cite{Tang2017OECC,Tanomura2020,Saygin2020,Tanomura2022PRA}. The unitary transformation can be almost any well-known N-mode mixer, which can significantly increase the flexibility and tolerance to fabrication errors.
However, configuring the phase shifters in this architecture is challenging, and no explicit configuration method has been known due to its complex mathematical structure.
The optimization of this architecture must deal with the many local minima present in its high-dimensional parameter space \cite{Saygin2020}. As a result, previous reports have relied on heuristic global searches, such as basin-hopping and simulated annealing, to configure the phase shifters \cite{Saygin2020, Tanomura2022PRA, Tanomura2021}. However, these methods are time-consuming and suffer from exponentially increasing search times as the parameter space dimension increases. To address this issue, a machine learning-based configuration algorithm has been proposed \cite{Kuzmin2021}. While this algorithm may offer a solution, it requires an accurate initial estimation of the structure and may result in decreased matrix fidelity if the initial estimation contains errors.

In this research, we present a new, fast and iteratively configurable MPLC architecture that does not require prior knowledge and relies only on output signals. This approach involves adding a few redundant layers to the existing MPLC architecture and using derivative-based optimization with gradient approximation. This additional layer redundancy significantly improves the optimization performance of the MPLC architecture, in contrast to the similar approach used for the Clements architecture \cite{Roel2017,Pai2019}, which adds a large number of redundant layers. When compared to numerical optimization of the Clements architecture without redundancy \cite{Pai2019}, our proposed method achieved 5 orders of magnitude better accuracy with 1/20 fewer iterations for $N=128$ modes of transformation. Additionally, our proposed method was able to achieve 5 orders of magnitude better accuracy and was 23 times faster in configuration compared to the previous report that used a heuristic algorithm to optimize the MPLC architecture \cite{Tanomura2020InP}.

This paper is structured as follows. Before discussing the main results, we begin by discussing important general properties of unitary matrix optimization and introducing a new distance in Section II. One key property we cover is that unitary matrix optimization essentially has no local minima.
Additionally, we propose a new distance, a phase-insensitive variant of the Frobenius norm, which is invariant under phase shifts at the output modes. Previously, the standard Frobenius norm has even been used in phase-insensitive applications.
In Section III, we investigate the optimization properties of the MPLC architecture with a few redundant layers of parametrization.
While the parametrization of a unitary matrix can cause optimization to fall into local minima, we demonstrate through numerical simulations that these can be effectively avoided by adding a few redundant layers. Our results show that this architecture can be efficiently optimized using well-known local minimization algorithms, such as the gradient descent algorithm, while the Clements architecture cannot. We also study the statistical properties of convergence.
In Section IV, we examine practical scenarios, such as when the gradient of the system is not available, only intensity detection is used, and crosstalk between phase shifters exists. We evaluate the impact of gradient approximation and crosstalk on the proposed method, and show that it still performs well, albeit with a reduction in achieved matrix accuracy after optimization or an increase in the number of iteration until convergence. The phase-insensitive distance exhibits similar optimization properties. In Section V, we conclude the paper.

\section{Matrix distance using the Frobenius norm}
This section presents some general mathematical properties of the Frobenius norm and proposes a new distance. We begin by defining the concept of unimodality for functions on the unitary group $U(N)$ and show that the matrix distance using the Frobenius norm exhibits this unimodality. We then clarify the range and expected value of the norm. Additionally, we introduce the phase-insensitive matrix distance for applications that only use intensity detection.

\subsection{Unimodality on $U(N)$}
Here, the concept of unimodality for a function on $U(N)$ is introduced. Unimodality is typically defined for probability distributions \cite{mathworldUnimodal}. For a multivariable function $f: \mathbb{R}^N \to \mathbb{R}$, unimodality is defined through the level set $L(f, \alpha)=\{ \bm{x}|f(\bm{x}) \leq \alpha, \bm{x} \in \mathbb{R}^N \}$ and the convexity of $L(f, \alpha)$ \cite{Anderson1955,Anescu2018}. In this paper, we extend this definition to functions on $U(N)$ by considering the path-connectedness of $L(f, \alpha)$, as $U(N)$ is not a convex set.

\textbf{Definition} A function $f:U(N) \to \mathbb{R}$ is called \textit{unimodal} if the level set $L(f, \alpha) = \{ X|f(X) \leq \alpha, X \in U(N) \} $ is path-connected for any $\alpha \in \mathbb{R}$.

In other words, any local minimum of a \textit{unimodal} function on $U(N)$ is also a global minimum.

\subsection{Unimodality of the Frobenius norm}
\label{sec:unimodalty_frobenius_norm}
We prove that the unitary matrix distance using the Frobenius norm is unimodal. The distance between two unitary matrices $d(X, U)$ is defined as $\norm{X-U}_F$, where $\norm{A}_F=\sqrt{\Tr\qty[A^\dag A]}$ is the Frobenius norm. It is worth noting that the matrix distance using mean square error (MSE) $\sum_{i,j} |X_{ij} - U_{ij}|^2$ is equivalent to $d(X, U)^2$. Given a unitary matrix $U \in U(N)$, we show that the function $f_U :U(N)\to\mathbb{R}$, defined as $f_U(X) = d(X, U)$ is unimodal. First, from the definition of the Frobenius norm, $f_U(X)^2$ is simplified as
\begin{equation}
\label{eq:f_U_intro}
\begin{split}
f_U(X)^2 &= \Tr\qty[(X-U)^\dag (X-U)] \\
&=2N-2\Re\qty[\Tr\qty[U^\dag X]].
\end{split}
\end{equation}
We write the eigenvalues of $U^\dag X$ as $\lambda_k (1 \leq k \leq N)$. Since both $U$ and $X$ are unitary, all the eigenvalues $\lambda_k$ satisfy $|\lambda_k|=1$. Therefore, the eigenvalues can be written as $\lambda_k = e^{i\theta_k}$, where $-\pi \leq \theta_k \leq \pi$. Using these eigenvalues, Eq. \ref{eq:f_U_intro} can be simplified further as
\begin{equation}
\label{eq:f_U_cos}
f_U(X)^2 = 2N-2\sum_{k=1}^N \cos \theta_k.
\end{equation}
Eq. \ref{eq:f_U_cos} implies that $f_U(X)^2$ is \textit{unimodal} because $\cos \theta_k$ is unimodal over the range $-\pi \leq \theta \leq \pi$ and their sum is also unimodal. An algebraic proof of this unimodality is provided in the Appendix \ref{sec:proof_unimodality}.

\subsection{Range and normalization}
\label{sec:range_and_normalization}
We derive the range of $f_U(X)^2$ from Eq. \ref{eq:f_U_cos} and propose a proper normalization for the distance. The maximum of $f_U(X)^2$ is $4N$ if and only if $\theta_k = \pm \pi$ for all $k$, and the minimum is $0$ if and only if $\theta_k = 0$ for all $k$. In previous studies, $f_U(X)^2$ has been normalized by $N$ \cite{Hamerly2022Nat}, $2N$ \cite{Pai2019}, or $N^2$ \cite{Tanomura2022PRA}. Here, we propose a normalization by $4N$, which yields $0 \leq f_U(X)^2/4N \leq 1$. This is a good normalization of the norm with a range from 0 to 1 that is independent of $N$.

\subsection{Expected value}
To calculate the expected value $\mathbb{E}\qty[f_U(X)^2/4N]$, the distribution of $\theta_k$ is considered. If $X$ is sampled from the Haar measure, then $U^\dag X$ is also Haar-random due to the invariance of the Haar measure. As a result, the eigenvalues of $U^\dag X$ are uniformly distributed on the unit circle $\abs{c}=1$, and we have $\theta_k \sim U(-\pi, \pi)$. Because
\begin{equation}
  \mathbb{E}\qty[\cos \theta_k] = \int_{-\pi}^{\pi} \frac{1}{2\pi} \cos \theta \,\mathrm{d\theta} = 0,
\end{equation}
the expected value of the second term in the Eq. \ref{eq:f_U_cos} is $0$. We now conclude that $\mathbb{E}\qty[f_U(X)^2 / 4N] = 2N / 4N = 0.5$. This fact is observed numerically in the initial value of the convergence plots in Sec. \ref{sec:few_layer_redundancy}.

\subsection{Phase-insensitive distance}
\label{sec:phase_insensitive_distance}
\begin{figure}[ht]
\begin{minipage}[b]{0.95\linewidth}
    \centering
    \includegraphics[width=5cm]{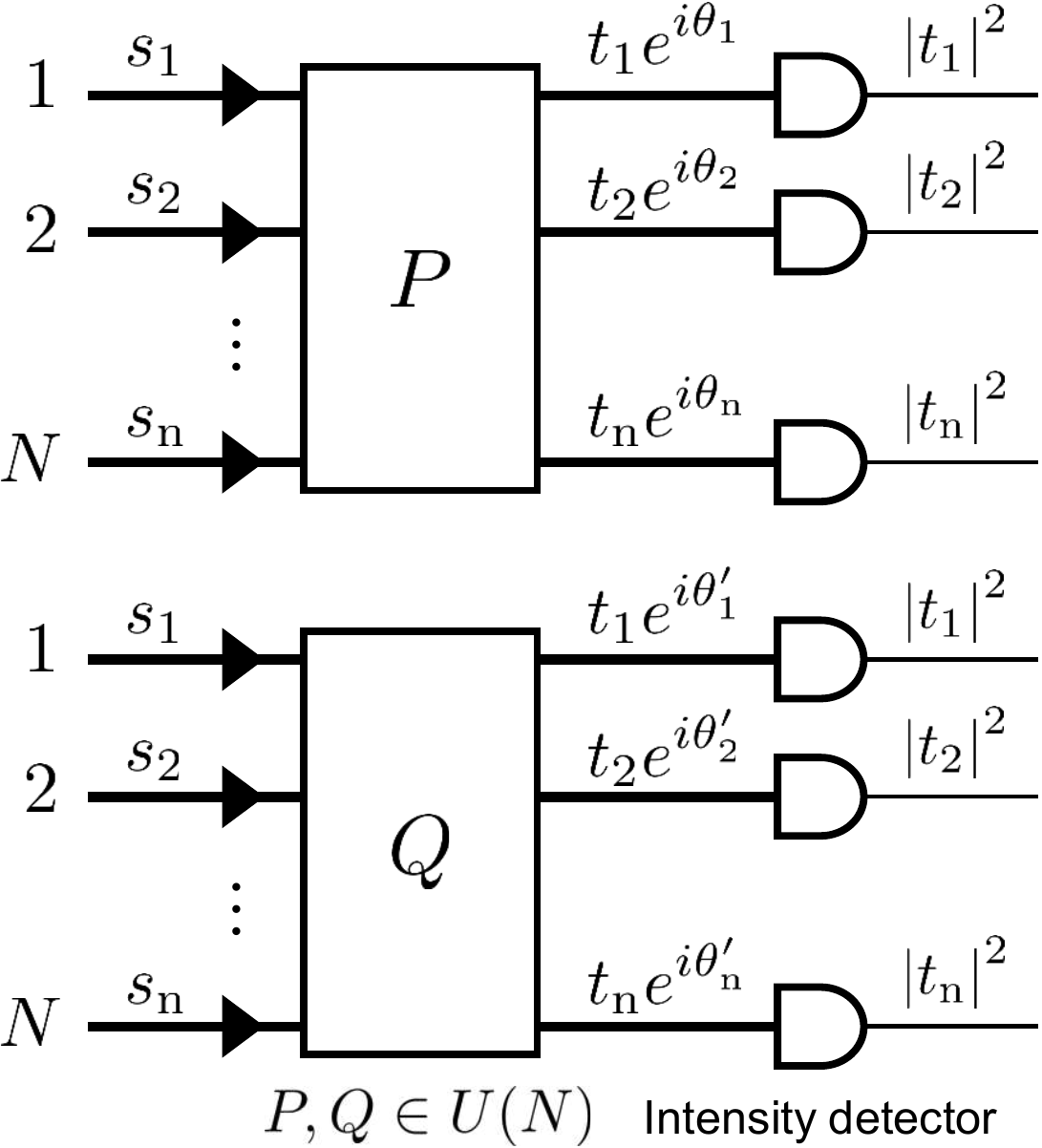}
\end{minipage}
  \caption{A complex vector $(s_1, s_2, \ldots, s_n)^\top$ is input into two unitary conversion devices, whose transfer matrices are represented as $P$ and $Q$. The outputs from these devices are identical, with the exception of the phase degrees of freedom at the outputs, when evaluated using a phase-insensitive distance.}
\label{fig:phase_insensitivity}
\end{figure}

Here, we introduce a phase-insensitive variant of the matrix distance using the Frobenius norm. This variant is suitable for applications that only detect the intensity of the output modes, as the distance should not be affected by the output phases from the unitary converter. Applications that benefit from this phase-insensitive distance include machine learning and quantum photonics, where photodiodes or photon number counters are placed at the output ports. Fig. \ref{fig:phase_insensitivity} shows a scenario where a complex vector $(s_1, s_2, \ldots, s_n)^\top$ is input into two unitary conversion devices. The transfer matrix for these devices is represented by $P$ and $Q$, and their complex outputs are in polar form as $t e^{i\theta}$. The only difference in the output vectors from these two devices is in their phase, with $\theta_i \neq \theta_i'$. In applications that only detect the intensity of output modes, these two matrices $P$ and $Q$ are treated the same and a suitable matrix distance is introduced for this purpose. In the following discussion, the matrix $U$ represents the given target unitary matrix, and the matrix $X$ represents the actual conversion achieved by the unitary converter device.

To investigate the effect of output phases from the unitary converter, we represent the unitary matrices $U^\dag$ and $X$ as
\begin{equation}
\label{eq:UX_intro}
\begin{gathered}
U^\dag = \begin{bmatrix}
\vb{u}_1 & \vb{u}_2 & \cdots & \vb{u}_n
\end{bmatrix}\\
X = \begin{bmatrix}
\vb{x}_1^\top \\[4pt]
\vb{x}_2^\top \\[4pt]
\vdots \\[4pt]
\vb{x}_n^\top
\end{bmatrix},
\end{gathered}
\end{equation}
where $\vb{u}_i$ are column vectors of $U^\dag$ and $\vb{x}_i^\top$ are row vectors of $X$. Since $U^\dag$ and $X$ are unitary matrices, the norms of $\vb{u}_i$ and $\vb{x}_i$ are all equal to 1. In this context, the output phases of the unitary converter correspond to the phases of $\vb{x}_i$. The original Frobenius norm changes its value under the global phase change of $\vb{x}_i$, which is defined as replacing $\vb{x}_i$ with $e^{i\theta_i}\vb{x}_i$.

We analyze the dependence of the Frobenius norm on the global phase of each $\vb{x}_i$. The Frobenius norm is expanded using the column vectors $\vb{u}_i$ as
\begin{equation}
\label{eq:f_U2_decomp}
\begin{split}
&\norm{X-U}^2_F = \norm{XU^\dag - I}^2_F \\
&= \norm{X\vb{u}_1-\vb{e}_1}^2 + \norm{X\vb{u}_2-\vb{e}_2}^2 + \cdots + \norm{X\vb{u}_n-\vb{e}_n}^2 ,
\end{split}
\end{equation}
where $\vb{e}_i$ is a unit column vector whose $i$-th element is $1$ and the others are $0$.
Expanding the term $\norm{X\vb{u}_1-\vb{e}_1}^2$, we obtain
\begin{equation}
\label{eq:Xu_e_norm}
\begin{split}
&\norm{X\vb{u}_1 - \vb{e}_1}^2 \\
&= |\vb{x}_1^\top \vb{u}_1 - 1|^2 + |\vb{x}_2^\top \vb{u}_2|^2+ \cdots + |\vb{x}_n^\top \vb{u}_n|^2 .
\end{split}
\end{equation}
Except for the first term $|\vb{x}_1^\top \vb{u}_1 - 1|^2$, the other terms $|\vb{x}^\top_i\vb{u}_i|^2 (i \geq 2)$ are invariant under the global phase change in $\vb{x}_i$. Only the first term depends on the global phase of $\vb{x}_1$.
Therefore, the term $\norm{X\vb{u}_1 - \vb{e}_1}^2$ is invariant under the global phase change in the row vectors $\vb{x}_2, \vb{x}_3, \cdots \vb{x}_n$. Similarly, $\norm{X\vb{u}_i - \vb{e}_i}^2$ is independent of the global phase of $\vb{x}_j$ where $j \neq i$.
When the global phase of $\vb{x}_i$ is changed, the term $|\vb{x}_i^\top \vb{u}_i - 1|^2$ takes its minimum value if and only if $\vb{x}^\top_i \vb{u}_i$ is a positive real number, because
\begin{equation}
\label{eq:Xu-1_sq}
|\vb{x}^\top_i\vb{u}_i - 1|^2 = |\vb{x}^\top_i \vb{u}_i|^2 - 2\Re\qty[\vb{x}^\top_i \vb{u}_i]+1
\end{equation}
and only the second term $2\Re\qty[\vb{x}^\top_i \vb{u}_i]$ is dependent on the global phase of $\vb{x}_i$. This motivates the idea of replacing all instances of $|\vb{x}_i^\top\vb{u}_i-1|^2$ with $\qty(|\vb{x}_i^\top\vb{u}_i|-1)^2$ in the Frobenius norm $\norm{X-U}^2_F$. As discussed in Eq. \ref{eq:Xu_e_norm} and Eq. \ref{eq:Xu-1_sq}, the minimum value of the Frobenius norm under this substitution is the same as the minimum value of the original Frobenius norm, due to the phase-dependence property.

Based on the aforementioned consideration, we propose a phase-insensitive matrix distance.
Given unitary matrix $U$, we define the distance function $h_U : U(N) \to \mathbb{R}$ using only terms whose form is $|\vb{x}_i^\top \vb{u}_j|$, which can be obtained using intensity measurements at the outputs from the unitary converter. We define positive real numbers $a_{ij}=|\vb{x}^\top_i \vb{u}_j|^2 \geq 0$.
All the $a_{ij}$ can be obtained by multiplying column vector $\vb{u}_i$ with matrix $X$ through the unitary converter, because
\begin{equation}
\label{eq:Xu_decomp}
X\vb{u}_j = \begin{bmatrix}
\vb{x}_1^\top \vb{u}_j \\[4pt]
\vb{x}_2^\top \vb{u}_j \\[4pt]
\vdots \\[4pt]
\vb{x}_n^\top \vb{u}_j
\end{bmatrix}
\end{equation}
and $|\vb{x}^\top_i \vb{u}_j|^2$ can be obtained through the intensity measurement of $\vb{x}^\top_i \vb{u}_j$. By expanding all the terms in Eq. \ref{eq:f_U2_decomp} with Eq. \ref{eq:Xu_e_norm} and applying the substitution discussed with respect to Eq. \ref{eq:Xu-1_sq}, we define $h_U(X)$ as
\begin{align}
\label{eq:xu_sq}
&\begin{alignedat}{4}
&h_U(X) = \\
&(\sqrt{a_{11}}-1)^2 & &+a_{12} & &+\cdots & &+ a_{1n} \\
&+a_{21} & &+(\sqrt{a_{22}}-1)^2 & &+\cdots & &+ a_{2n} \\
&\vdots & &\vdots & &+(\sqrt{a_{ii}}-1)^2 & & \cdots\\
&+a_{n1} & &+a_{n2} & &+\cdots & &+(\sqrt{a_{nn}}-1)^2 \\
\end{alignedat} \nonumber \\
  &=\sum_{ij} \qty(\delta_{ij} - \abs{\qty[XU^\dag ]_{ij}})^2.
\end{align}
This function is independent of the global phase of $\vb{x}_i$, that is, the phases of each output from the unitary converter, and has a minimum value identical to the original Frobenius norm distance function $f_U(X)$.

The function $h_U(X)$ is also unimodal. This can be proven by reductio ad absurdum. Suppose $h_U(X)$ is not a unimodal function and has multiple disconnected local minima. Then, for all matrices $X'$ such that $h_U(X')$ is a local minimum, there must exist at least one set of phases $(\theta_1, \theta_2, \cdots \theta_n)$ that defines a matrix
\begin{equation}
Y = \begin{bmatrix}
e^{i\theta_1} \vb{x}_1'^\top \\[4pt]
e^{i\theta_2} \vb{x}_2'^\top \\[4pt]
\vdots \\[4pt]
e^{i\theta_n} \vb{x}_n'^\top
\end{bmatrix}
\end{equation}
such that $f_U(Y)$ is also a local minimum, where $\vb{x}_i'^\top$ is the row vector of $X'$. The existence of such a matrix $Y$ follows from the definition of $h_U(X)$. However, this would mean that $f_U(X)$ also has multiple disconnected local minima, which is a contradiction. Therefore, $h_U(X)$ must be unimodal.

\section{Few-layer redundant parameterization}
\label{sec:few_layer_redundancy}
The global optimization property guaranteed by unimodality discussed in the previous section is derived without any assumptions about the matrix being optimized; however, the matrix synthesized by the device is parametrized by physical parameters and unimodality may not always hold in this parameter space. We examine this issue and propose a solution to mitigate this difficulty.
Let $l_U(X) \geq 0$ be a distance function from a desired unitary matrix $U$, and $X(\vb{p})$ be a unitary matrix realized by a physical converter that is parametrized by a real parameter vector $\vb{p}$. Each element in $\vb{p}$ corresponds to the amount of phase shift in the actual device. If $l_U(X)$ is unimodal, its gradient becomes a zero vector only when $X=\pm U$. However, when optimizing an actual unitary conversion device, we need to consider the scalar optimization of $l_U(X(\vb{p}))$. If the Jacobian of $X(\vb{p})$ is full-rank at any $\vb{p}$, meaning there exists infinitesimal parameter changes $\Delta \vb{p}$ for any infinitesimal matrix changes $\Delta X$, then the function $l_U(X(\vb{p}))$ also has a single minimum due to the aforementioned unimodal property of $l_U(X)$. By increasing the number of layers in the unitary converter device, the degree of freedom in the parameter space increases, which may make the Jacobian of $X(\vb{p})$ more likely to be full-rank.
In this section, we demonstrate that increasing the number of layers in unitary converter devices by a few from its minimum requirement significantly improves the optimization of MPLC architecture using a gradient-based optimization algorithm.
\subsection{Device definition and redundancy}
\begin{figure}[t]
\begin{minipage}[b]{0.95\linewidth}
    \centering
    \vspace{1cm}
    \includegraphics[width=7cm]{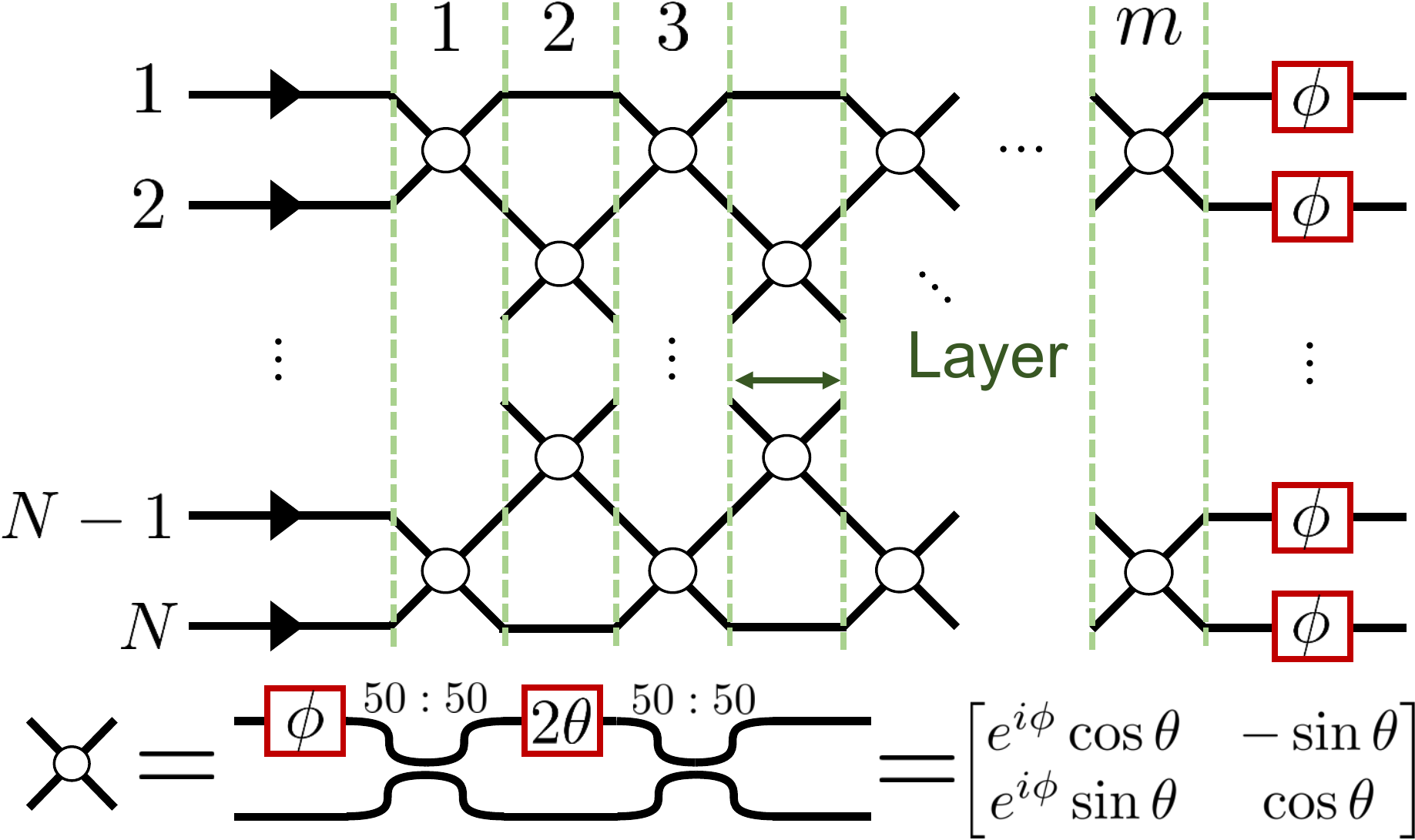}
    \subcaption{}
    \label{fig:schem_clements}
  \end{minipage}
  \begin{minipage}[b]{0.95\linewidth}
    \centering
    \vspace{0.5cm}
    \includegraphics[width=7cm]{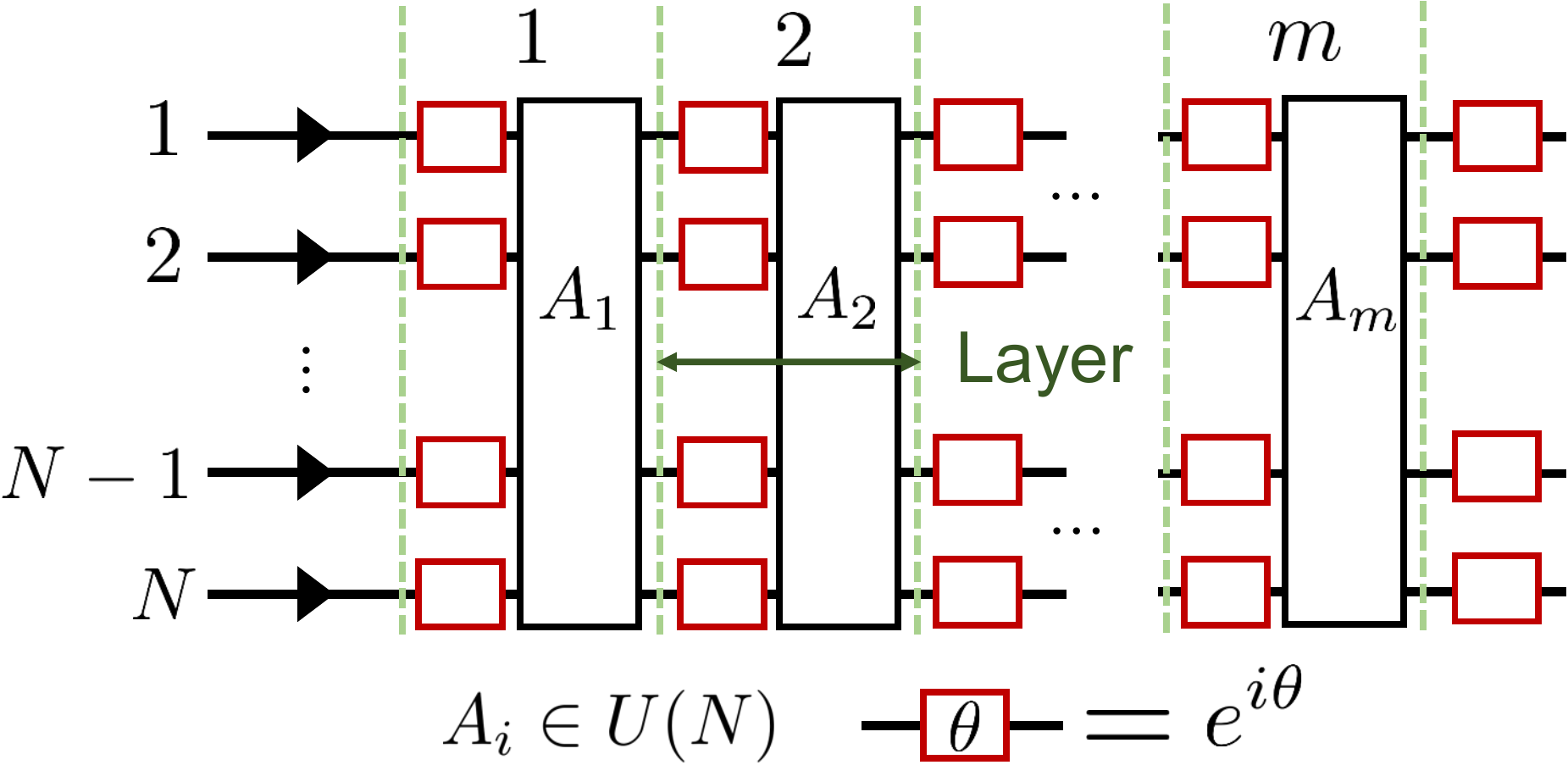}
    \subcaption{}
    \label{fig:schem_mplc}
\end{minipage}
  \caption{Schematics of the $N \times m$ Clements architecture (a) and $N \times m$ MPLC architecture (b). The left ports are inputs, and the right ports are outputs. The number of layers in each architecture is specified by $m$. In the Clements architecture, each layer contains either $N/2$ or $(N-1)/2$ MZI nodes, represented by white circles in the figure. Each MZI node consists of two phase shifters, represented by the variables $\phi$ and $\theta$. In the MPLC-based unitary converter (b), the architecture consists of an $N$-port fixed unitary converter represented by $A$, followed by an array of $N$ single-mode phase shifters.}
\label{fig:schem_unitary_converters}
\end{figure}

We present the mathematical definition of the unitary converters and the few-layer redundant parameterization. Fig. \ref{fig:schem_clements} shows the architecture of the MZI-based unitary converter, which is commonly referred to as Clements architecture \cite{Clements2016}. The MZI consists of two $50:50$ BSs and two phase shifters, which can realize an arbitrary $U(2)$ transformation. In this paper, we do not consider any imperfections of the MZI.
Fig. \ref{fig:schem_mplc} shows the structure of the MPLC architecture. Each layer consists of an $N$-port fixed unitary converter $A_i$ and an array of $N$ single-mode phase shifters. After $m$ layers, another array of phase shifters is placed in a similar manner to the Clements architecture. The overall transformation of this device, denoted as $X$, is given by
\begin{equation}
X = L_{m+1} A_m L_m \cdots A_2 L_2 A_1 L_1,
\end{equation}
where $A_i$ is the transfer matrix of a $N$-port unitary converter and $L_i$ is expressed as
\begin{equation}
L_i = \begin{bmatrix}
e^{i\theta_{i1}} & & & \\
& e^{i\theta_{i2}} & & \\
& & \ddots & \\
& & & e^{i\theta_{in}}
\end{bmatrix}.
\end{equation} For any $i \neq j$, the matrices $A_i$ and $A_j$ are different. The total number of degrees of freedom in this matrix is $(m+1)(N-1)+1$. This is because each phase shifter array has $N-1$ degree of freedom due to the loss of one degree of freedom from the global phase, and the entire device has an additional degree of freedom, the global phase. The $N$-port fixed unitary converter $A_i$ can be implemented using a multiport directional coupler \cite{Tanomura2020}, multimode interference coupler \cite{Tang2017}, or other multiport unitary transform devices. The device should be carefully chosen to ensure that the overall transformation $X$ is universal. The mixing entropy of a device can be used as a measure of universality \cite{Tang2021, Tanomura2022}. To realize an arbitrary $U(N)$ transformation, the total number of degrees of freedom must exceed $N^2$ \cite{Reck1994}. For the Clements architecture, the number of layers $m$ must satisfy $m \geq N$ \cite{Clements2016}. Similarly, the number of layers $m$ for the MPLC architecture must also satisfy $m \geq N$, which follows from $(m+1)(N-1)+1 \geq N^2$. In this context, a few-layer redundant parameterized architecture is defined as having $m=N+1, N+2$ layers for both architectures.

\begin{figure*}[htbp]
    \begin{tabular}{cc}
      \begin{minipage}[t]{0.48\hsize}
        \centering
        \includegraphics[width=8cm]{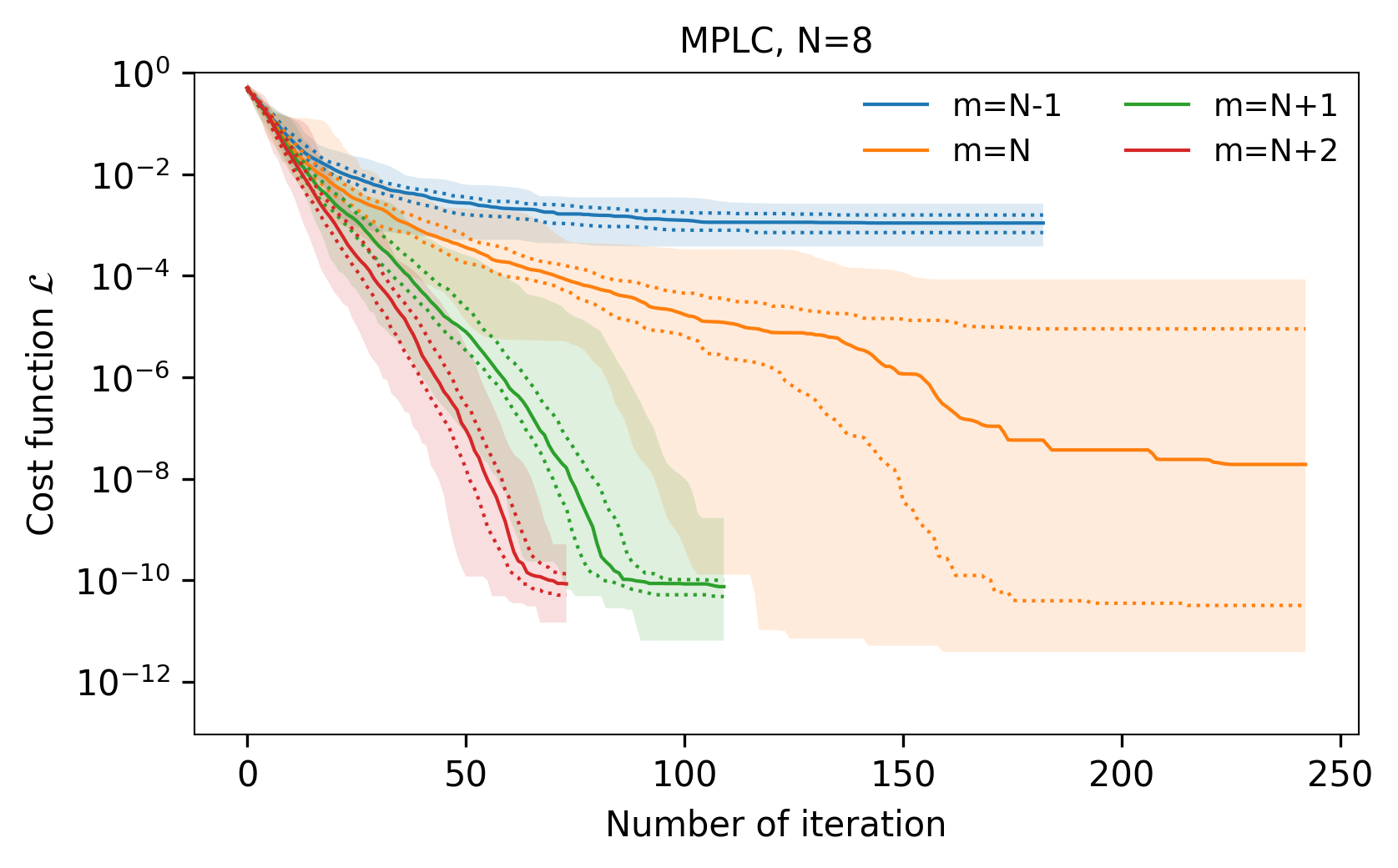}
        \subcaption{}
        \label{fig:LBFGS_comp_MPLCN8}
      \end{minipage} &
      \begin{minipage}[t]{0.48\hsize}
        \centering
        \includegraphics[width=8cm]{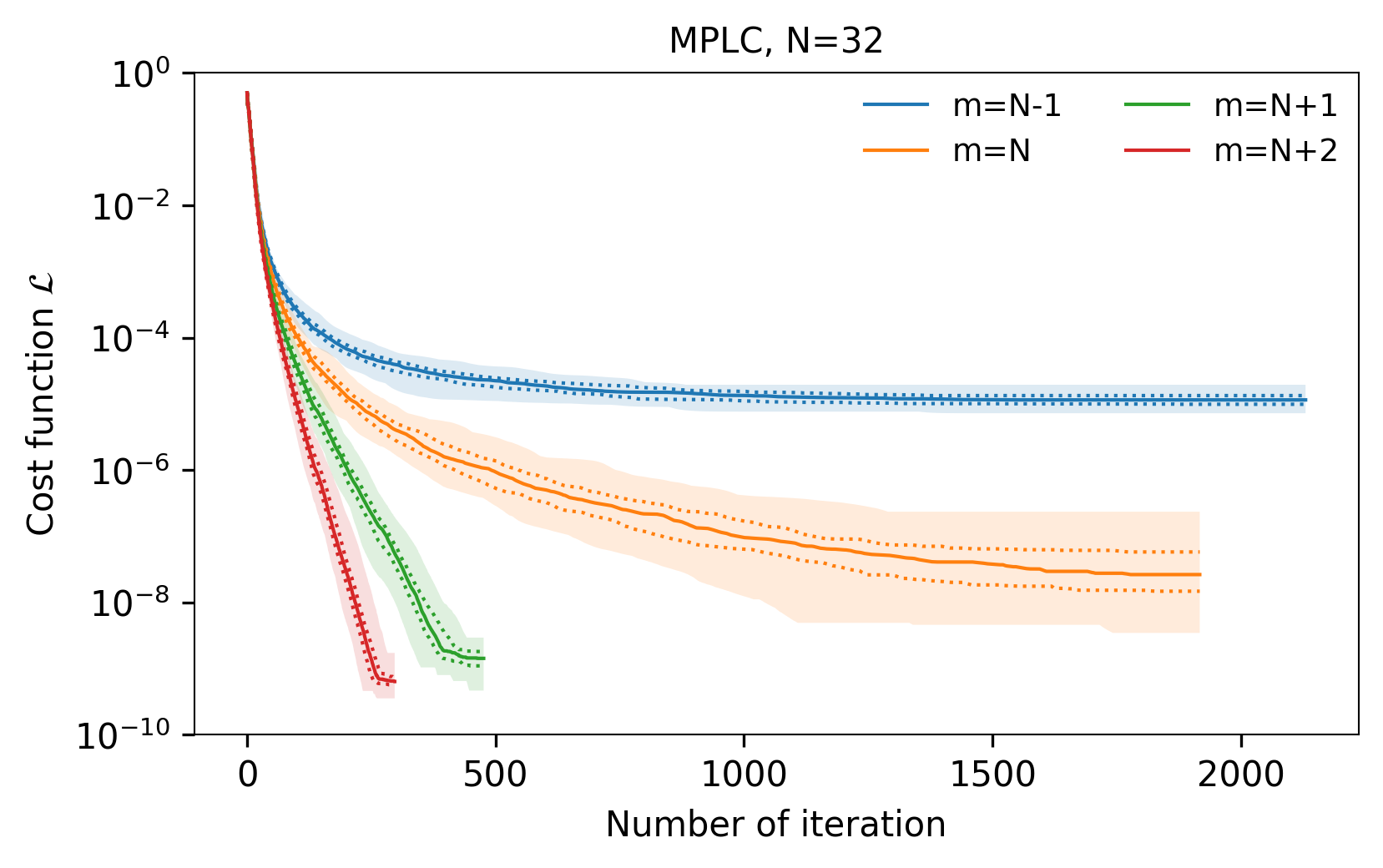}
        \subcaption{}
        \label{fig:LBFGS_comp_MPLCN32}
      \end{minipage} \\

      \begin{minipage}[t]{0.48\hsize}
        \centering
        \includegraphics[width=8cm]{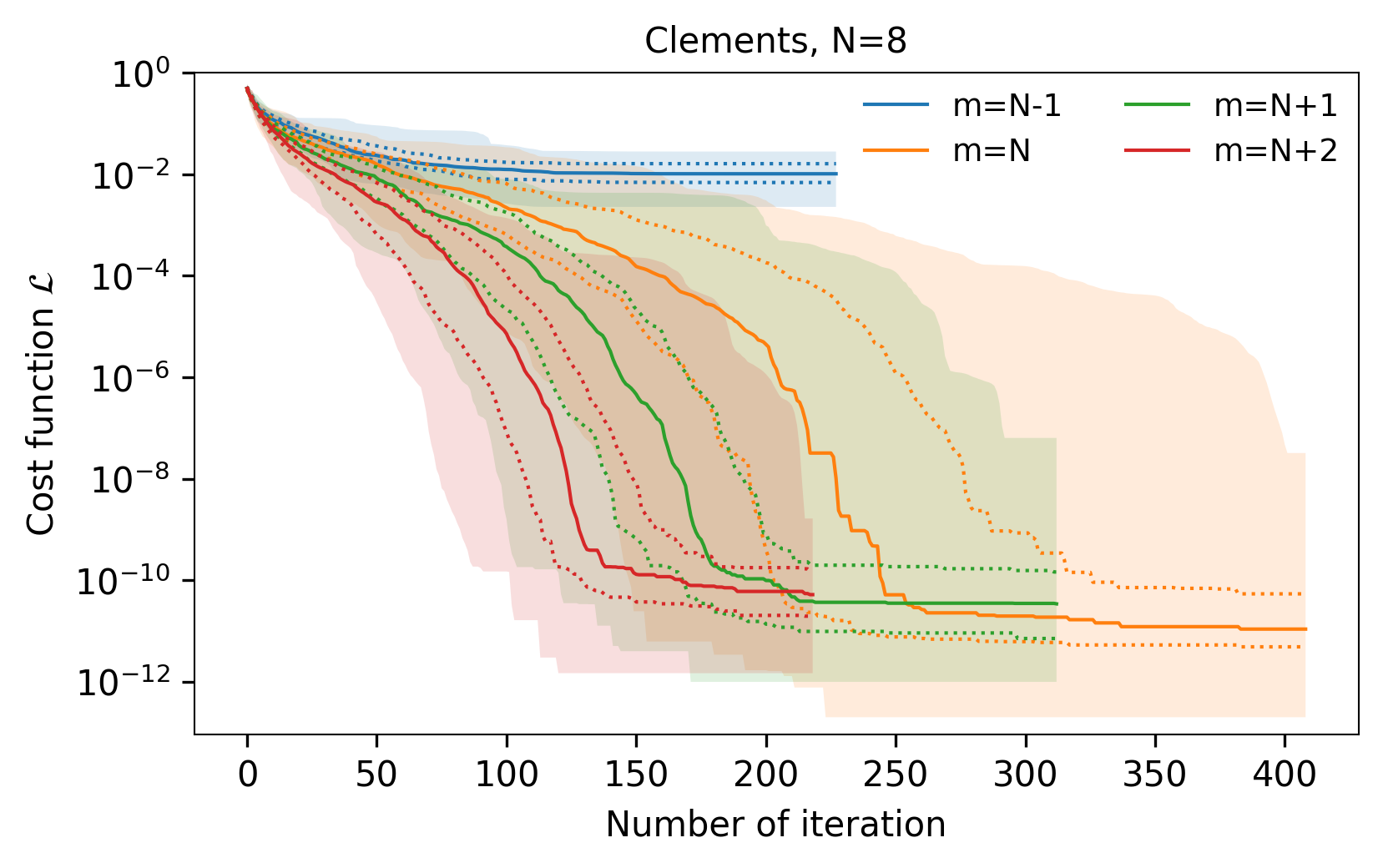}
        \subcaption{}
        \label{fig:LBFGS_comp_ClementsN8}
      \end{minipage} &
      \begin{minipage}[t]{0.48\hsize}
        \centering
        \includegraphics[width=8cm]{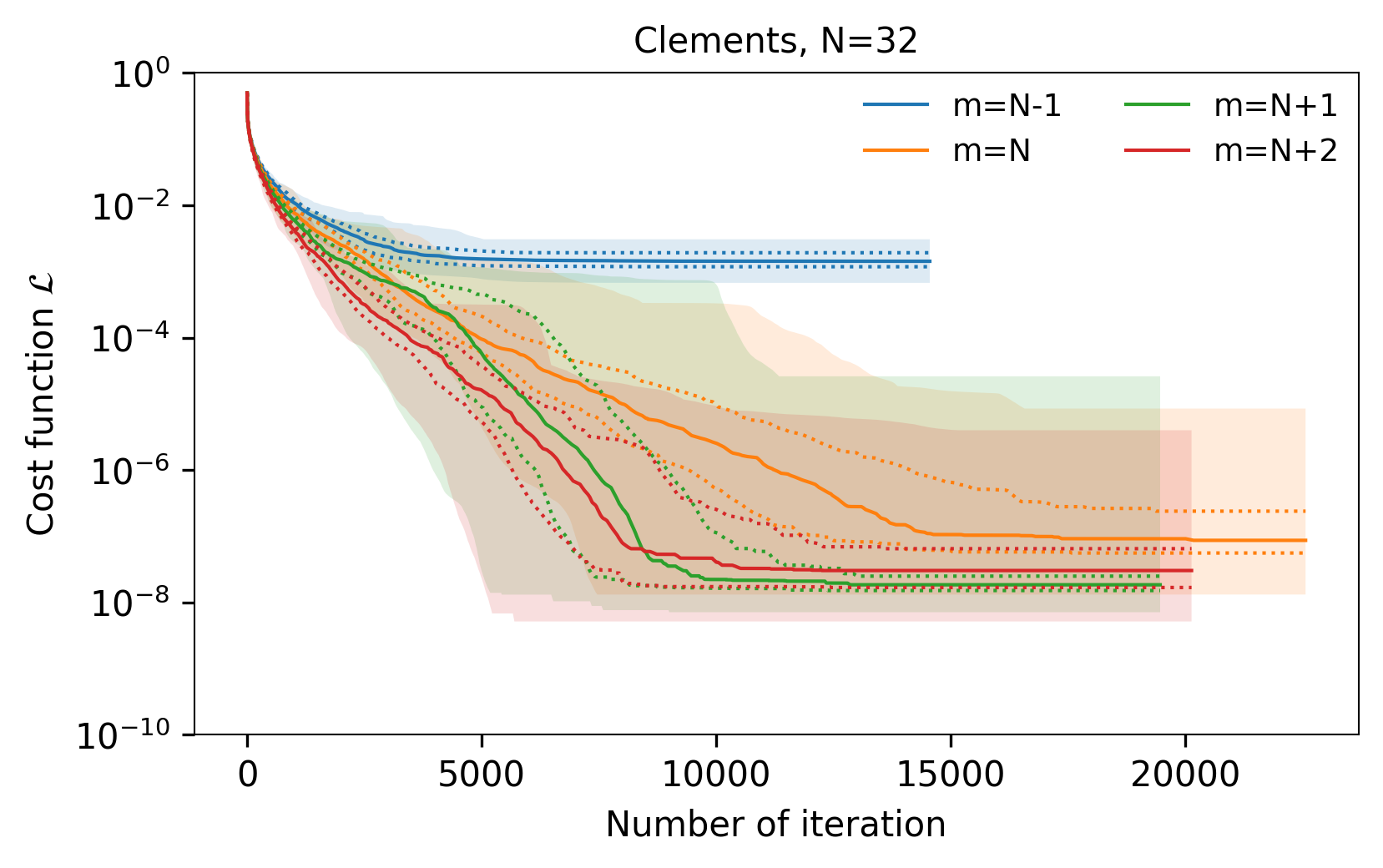}
        \subcaption{}
        \label{fig:LBFGS_comp_ClementsN32}
      \end{minipage}
    \end{tabular}
    \caption{Convergence plots for MPLC and Clements architectures. The vertical axis shows the value of the cost function $\mathcal{L}$ defined by Eq. \ref{eq:def_L}, and the horizontal axis shows the number of iterations. The shaded area represents the minimum and maximum values, the solid line represents the median, and the dotted line represents the 25\% and 75\% quantiles over 64 optimization trials. MPLC architecture with (a) $N=8$, (b) $N=32$, Clements architecture with (c) $N=8$, and (d) $N=32$.}
    \label{fig:LBFGS_comp}
\end{figure*}

\subsection{Optimization problem setting and algorithm}
\label{sec:optimization_problem_setting_and_algorithm}
We formulate the matrix optimization problem as follows. We have real parameter variables expressed as a vector $\vb{p}$. The number of parameters depends on $m$ and $N$. We define the normalized cost function $\mathcal{L}$ between two matrices as
\begin{equation}
\mathcal{L}(\vb{p}) = \frac{1}{4N} \norm{X(\vb{p}) - U}_F^2,
\label{eq:def_L}
\end{equation}
where $X({\vb{p}})$ is the unitary matrix realized physically by the parameter vector $\vb{p}$, $U$ is the target matrix to be achieved, and $\norm{\cdot}_F$ is the Frobenius norm.
The cost function $\mathcal{L}$ is divided by $4N$ as discussed in the Section \ref{sec:range_and_normalization}, and $0 \leq \mathcal{L} \leq 1$ is always satisfied. At the beginning of the optimization, parameters are initialized using uniform distribution ranging from $0$ to $2\pi$, and the target unitary matrix $U$ is sampled from Haar measure using the \texttt{stats} module of SciPy \cite{ScipyNmeth}. For the MPLC architecture, the matrix $A_i$ for $1 \leq i \leq m$ is also sampled from Haar measure. After initializing the parameters and matrices, the cost function $\mathcal{L}$ is optimized using the quasi-Newton optimization method L-BFGS \cite{Flet1987} implemented in \texttt{optimize} module of SciPy \cite{ScipyNmeth}.
The derivative of the cost function $\mathcal{L}$ required for L-BFGS is calculated using automatic differentiation with the JAX framework \cite{jax2018github}. This method starts from the initial parameters and modifies them at each step until convergence to the local minimum, where $d\mathcal{L}/d\vb{p}=\vb{0}$. In each layer, we have $N$ real parameters to represent $N$ phase shifts in both Clements and MPLC architectures. The optimization is run 64 times while changing the initial parameters to investigate the statistical behavior. Cases with $N=8$ and $N=32$ are investigated.

\subsection{Results}
\label{sec:redundant_param_res}
Fig. \ref{fig:LBFGS_comp} shows the convergence plots when the number of layers is changed. The convergence plot of the cost function is recorded for 64 optimization trials. The shaded area shows the range of minimum and maximum values, the dotted line shows the 25\% and 75\% quantiles, and the solid line shows the median of the trials. For both Clements and MPLC architecture, the insufficient parameterized layer setting results in a large amount of errors. For MPLC architecture, the non-redundant case of $m=N$ results in a large variance of error, especially for $N=8$. This suggests the presence of many local minima in the parameter space of the MPLC architecture, as previously reported in Ref. \cite{Saygin2020}. Although the variance of non-redundant setting $m=N$ of $N=32$ is smaller than that of $N=8$, the error still remains for $N=32$, indicating the presence of inevitable local minima for this condition as well. When we increase the number of layers and add redundant degrees of freedom, the variance and error become small, as shown in the cases of $m=N+1, N+2$. In contrast, the Clements architecture results in a large variance of error for all conditions, even though it has the sufficient number of degrees of freedom.

\begin{figure}[ht]
\begin{minipage}[b]{0.95\linewidth}
    \centering
    \includegraphics[width=8cm]{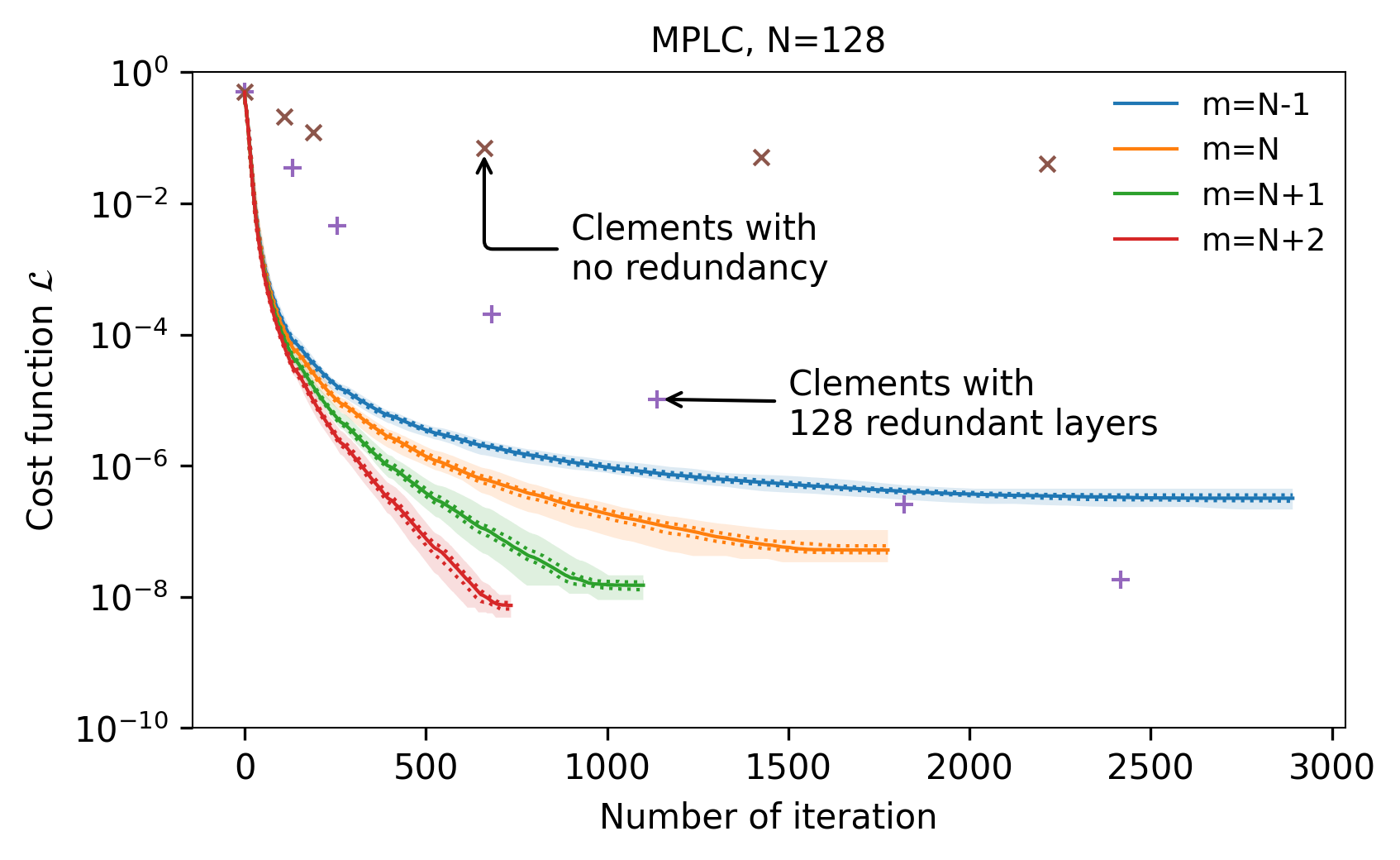}
\end{minipage}
  \caption{Comparison of the large case of $N=128$ with the previous study \cite{Pai2019}.}
\label{fig:large_comp_N_128}
\end{figure}

For the cases with a large number of ports, $N=128$, Fig. \ref{fig:large_comp_N_128} shows the performance comparison with the previous study \cite{Pai2019} of the Clements architecture with redundancy. When compared with no redundancy, the previous study converges at $\mathcal{L} = 1.4\times10^{-2}$ after 20000 iterations, while MPLC architecture yields a result that is 6 orders of magnitude better than the Clements architecture with 1/20 fewer iterations. The MPLC architecture with a redundant layer $m=N+1$ still outperforms in terms of both convergence speed and accuracy, even compared with the Clements architecture with 128 redundant layers.

\begin{figure}[hbtp]
\begin{minipage}[b]{0.49\linewidth}
    \centering
    \includegraphics[width=4.3cm]{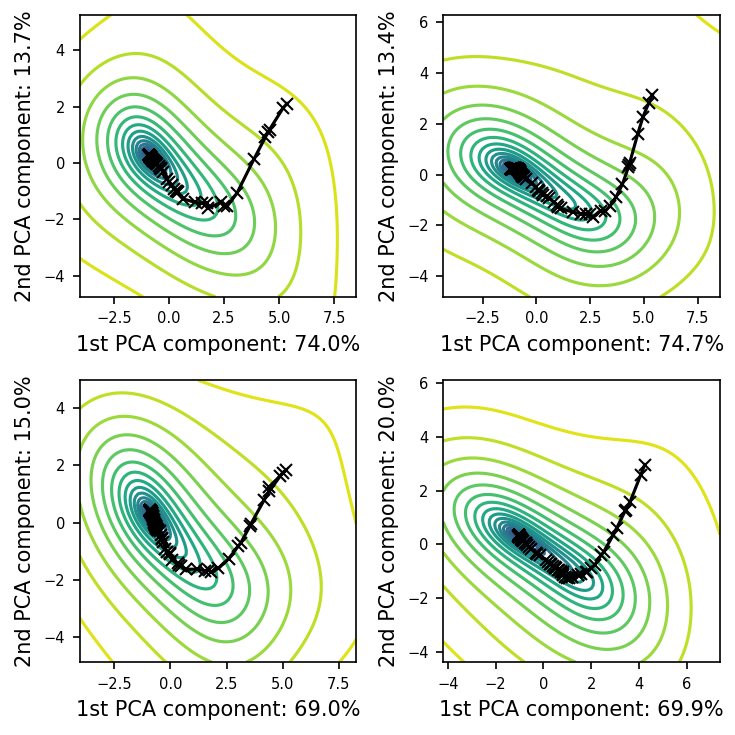}
    \subcaption{}
    \label{fig:MPLC_landscape}
  \end{minipage}
  \begin{minipage}[b]{0.49\linewidth}
    \centering
    \includegraphics[width=4.3cm]{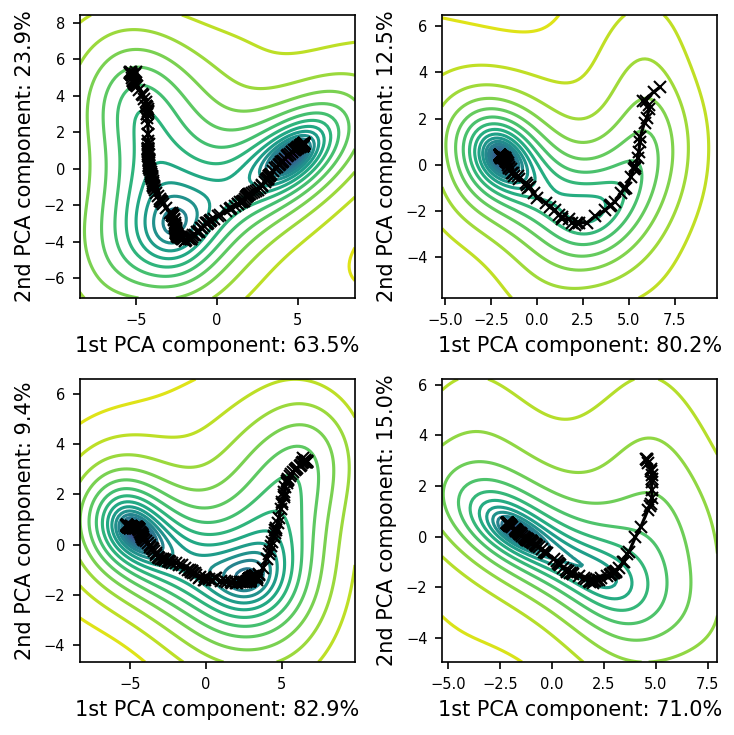}
    \subcaption{}
    \label{fig:clements_landscape}
\end{minipage}
  \caption{The PCA projections of the optimization trajectories and contour plot of the log of the loss function for the MPLC architecture (a) and the Clements architecture (b) for $N=8, m=N+1$. Each of the four figures shows the trajectory of the optimization when the initial parameters and the target matrix are randomly changed.}
\label{fig:optimization_landscape}
\end{figure}

We visualized the optimization trajectory and loss function in the parameter space using the method reported in \cite{Gallagher2003,Li2017} and in the supplementary material of \cite{Li2018}. The optimization trajectory is the path of the parameters in a high-dimensional space created by the optimization. We stored the parameter history at each step of the optimization and applied principal component analysis (PCA) to that history.
The first and second PCA components were used to project the high-dimensional path onto a two-dimensional space. The visualization was performed for $N=8, m=N+1$. Fig. \ref{fig:optimization_landscape} shows the projected trajectories and contour plots of the log of the loss function in the projected subspace. The contour plot for the MPLC architecture is like a simple elliptic unimodal function, while that of the Clements architecture is more complex. The difference in the contour plots between these architectures suggests the reason for the convergence plot difference shown in Fig. \ref{fig:LBFGS_comp}.

\section{Optimization under practical settings}
We discuss three challenges that must be addressed when applying the proposed method in Section \ref{sec:few_layer_redundancy} to real device optimization.
First, the optimization method used in this section requires the gradient of the target function. While the gradient can be taken physically \cite{Hughes2018}, it requires additional external equipment, which makes the system bulky and not scalable. Second, the optimization method uses the complex amplitudes at the output for optimization. Reading the complex amplitudes in a real device requires coherent detectors at the output, which complicates the device. While some applications require coherent detection, many photonics-based optical computing platforms and quantum computing with photonic chips use intensity detection. Third, real devices have crosstalk between phase shifters, which is not considered in the optimization method. Crosstalk is especially problematic in thermo-optic phase shifters \cite{Jacques2019,Gurses2022}, although they are attractive due to their small footprints.

In this section, we report the results of derivative-free optimization with the original and phase-insensitive norm, using only the output signal, and examine the effect of crosstalk. We first introduce the mathematical formulation and then present the numerical results. The optimization method used is the same as in Section \ref{sec:optimization_problem_setting_and_algorithm}. These results pave the way for the design of optical unitary converters without the need for additional components, making the platform more scalable and versatile.

\subsection{Gradient approximation of multivariate function}
We use numerical gradient approximation, which uses only function values to approximate the analytical gradient, and investigate the effect of this approximation on the optimization behavior. The gradient of multivariate function $\nabla f(x_1, x_2, \ldots, x_n)$ is approximated by
\begin{equation}
\nabla f(x_1, x_2, \ldots, x_n) \approx \begin{bmatrix}
  \frac{f(x_1 + \Delta, x_2, \ldots, x_n) - f(x_1, \ldots, x_n)}{\Delta} \\
  \frac{f(x_1, x_2 + \Delta, \ldots, x_n) - f(x_1, \ldots, x_n)}{\Delta} \\
  \vdots \\
  \frac{f(x_1, x_2, \ldots, x_n + \Delta) - f(x_1, \ldots, x_n)}{\Delta} \\
\end{bmatrix}.
\end{equation} where $\Delta \ll 1$ represents a finite difference. Calculating gradient approximation requires the same number of function evaluations as the number of parameters. We show that a derivative-based algorithm using such gradient approximation can still be effective for optimizing unitary matrices.

\subsection{Definition of device with intensity detection}
\begin{figure}[ht]
\begin{minipage}[b]{0.95\linewidth}
    \centering
    \includegraphics[width=7.5cm]{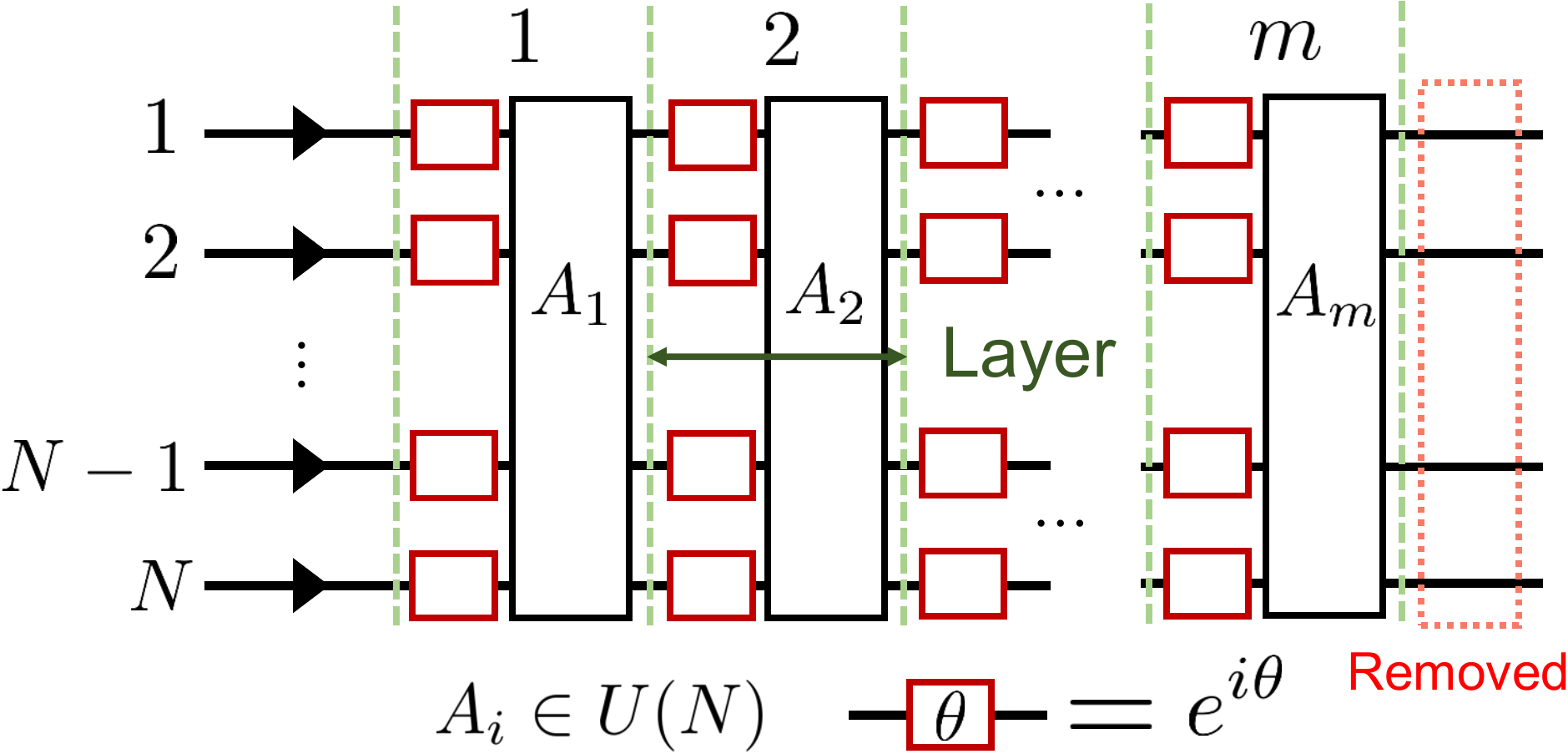}
\end{minipage}
\caption{Schematics of the MPLC architecture with the last phase shifter array removed.}
\label{fig:schem_MPLC_last_phase_removed}
\end{figure}

We evaluate the optimization behavior of the phase-insensitive distance introduced in Section \ref{sec:phase_insensitive_distance}, which only uses intensity detectors at the outputs. We expect the phase-insensitive distance to behave similarly to the phase-sensitive distance during optimization because it also has a unimodal property, as shown in Section \ref{sec:phase_insensitive_distance}. In order to test this, we removed the last phase shifter array from the MPLC architecture, as shown in Figure \ref{fig:schem_MPLC_last_phase_removed}. Although this removes the $N$ degree of freedom from the architecture, we still expect the optimization behavior to be similar to that of a standard phase-sensitive norm.

\subsection{Model of crosstalk}
\begin{figure}[htb]
    \centering
    \includegraphics[width=3cm]{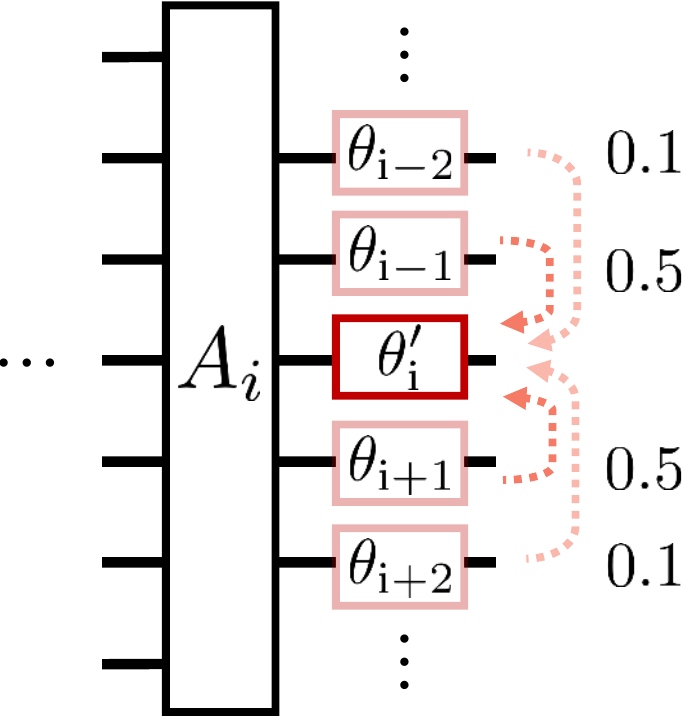}
  \caption{Crosstalk model of the MPLC architecture.}
\label{fig:crosstalk_scheme}
\end{figure}

We model crosstalk by considering the interaction between adjacent phase shifters. The crosstalk is represented by a linear combination of phase shifts, which can be expressed as
\begin{equation}
\theta_i = \sum_j \alpha_{ij} \theta_j.
\end{equation}
The coupling model and coupling coefficients $\alpha_{ij}$ are shown schematically in Fig. \ref{fig:crosstalk_scheme}. The coupling in the following simulations is formulated as $\theta_i' = 0.1\theta_{i-2} + 0.5\theta_{i-1} +\theta_{i} + 0.5\theta_{i+1} + 0.1\theta_{i+2}$.
If the coupled parameter $\vb{p}' = w(\vb{p})$ is reversible, the unitary matrix $X(\vb{p}')$ realized by the coupled parameter will have a full-rank Jacobian if the original $X(\vb{p})$ has a full-rank Jacobian. We use the gradient approximation for optimization.

\subsection{Results}
\begin{figure}[bhtp]
\centering
\includegraphics[width=8.5cm]{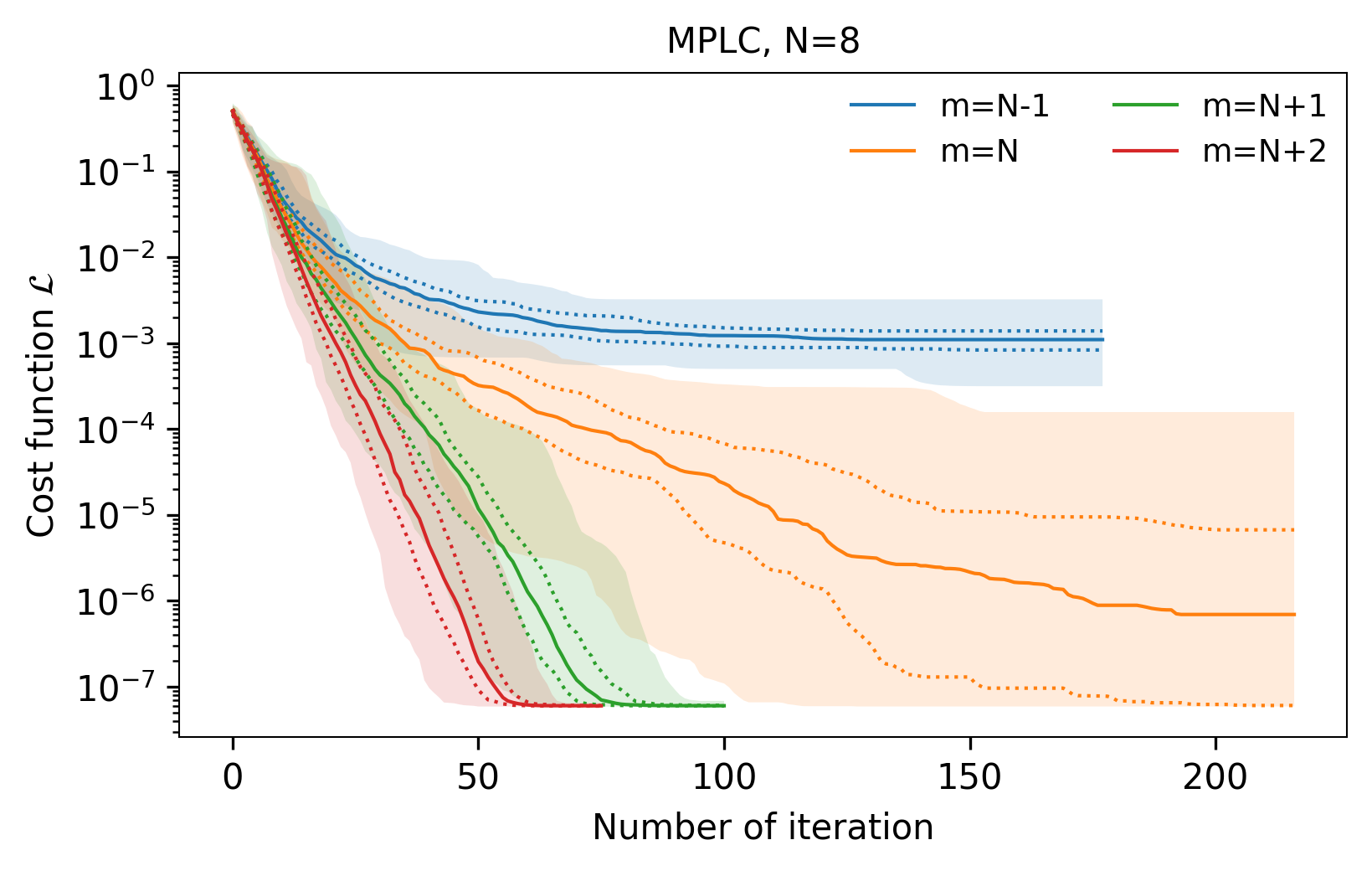}
\label{fig:approxgrad_comp_MPLCN8}
\caption{Convergence plots for the MPLC architecture with gradient approximation, where $\Delta = 2^{-10}$. The 64 optimization trials are shown in the same manner as in Fig. \ref{fig:LBFGS_comp}.}
\label{fig:approxgrad_comp_structure}
\end{figure}

Fig. \ref{fig:approxgrad_comp_structure} shows the convergence plots for gradient approximation using the standard Frobenius-norm-based distance. The approximation is calculated using a $\Delta$ value of $2^{-10}$, which corresponds to phase shifts with 10-bit resolution. When a redundant layer is added, the MPLC architecture shows numerical-accuracy limited performance (for $m=N+1, N+2$) with a small variance in the error. Each iteration of the optimization requires the evaluation of the distance the same number of times as the number of parameters due to the gradient approximation. For example, when $N=8$ and $m=N+1$, each optimization requires $8\times(8+2)=80$ evaluations of the distance. Using the MPLC architecture, the optimization converges at 100 iterations for this case, so the total number of evaluations is approximately 8000.

\begin{figure}[ht]
\begin{minipage}[b]{0.95\linewidth}
    \centering
    \includegraphics[width=8.5cm]{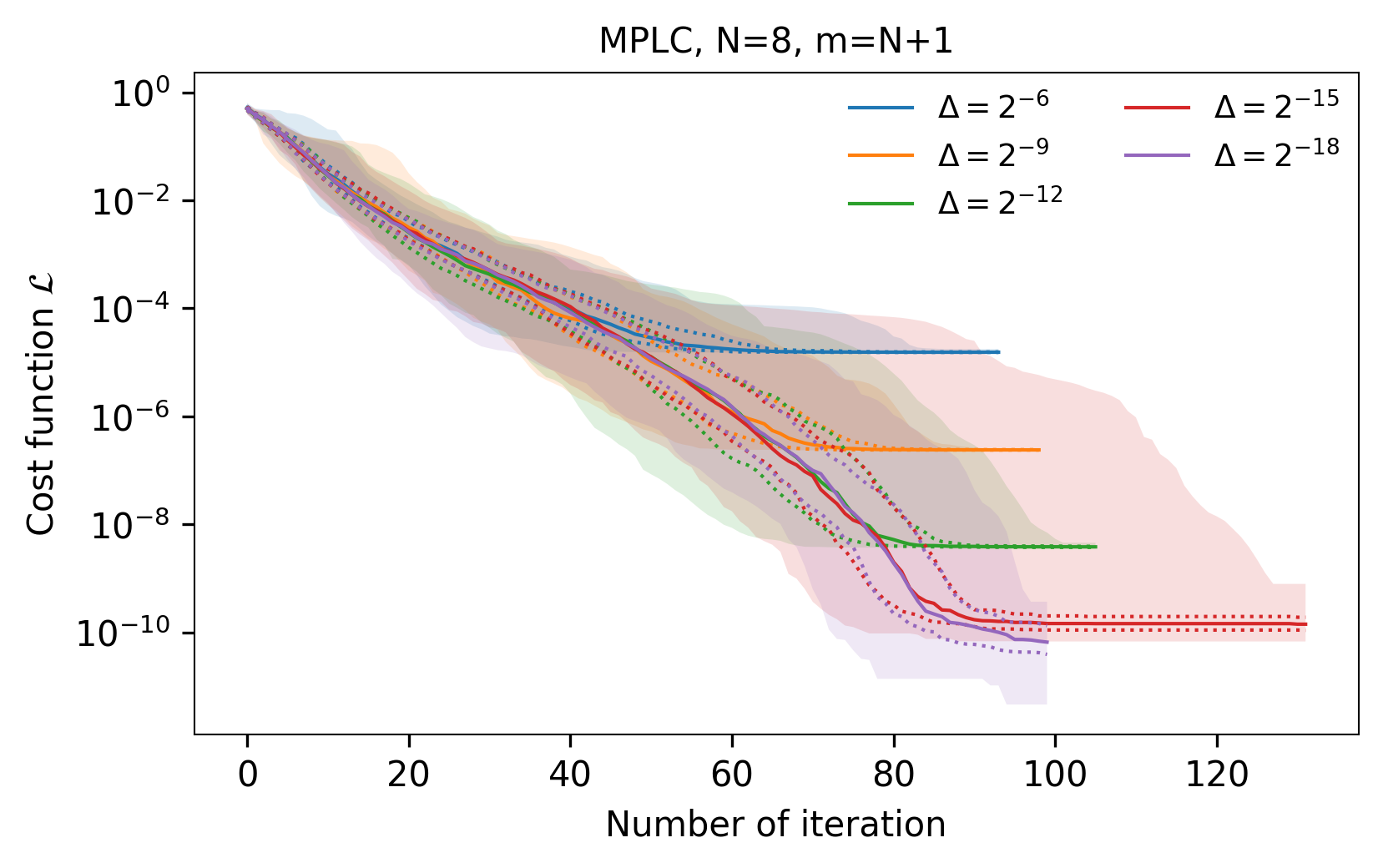}
\end{minipage}
  \caption{Comparison of final error when changing the accuracy of gradient approximation for the MPLC architecture. The 64 optimization trials are shown in the same manner as in Fig. \ref{fig:LBFGS_comp}.}
\label{fig:approx_grad_acc_comp_MPLC}
\end{figure}

We examined the $\Delta$ dependence of the small error variance observed in Fig. \ref{fig:approxgrad_comp_structure}, which arises from the gradient approximation. Fig. \ref{fig:approx_grad_acc_comp_MPLC} shows the optimization results for each gradient approximation accuracy using a redundant layer setting of $m=N+1$. As the finite difference $\Delta$ becomes smaller, the final error also becomes smaller. If the accuracy of the gradient approximation is not sufficient, meaning $\Delta$ is not small enough, the variance of the optimization result is very small.
For example, the error is in the range $\qty[1.5\times10^{-5}, 1.8\times10^{-5}]$ for the $\Delta = 2^{-6}$ case, $\qty[2.3\times10^{-7}, 2.5\times10^{-7}]$ for the $\Delta = 2^{-9}$ case, and $\qty[3.5\times10^{-9}, 4.5\times10^{-9}]$ for the $\Delta = 2^{-12}$ case.
If the accuracy of the gradient approximation is sufficient ($\Delta \le 2^{-15}$), the result has a non-negligible variance similar to the one shown in Fig. \ref{fig:LBFGS_comp_MPLCN8}. The error is in the range $\qty[7.9\times10^{-10}, 6.8\times10^{-11}]$ for the $\Delta = 2^{-15}$ case and $\qty[3.7\times10^{-10}, 4.6\times10^{-12}]$ for the $\Delta = 2^{-18}$ case.
These result provide criteria for designing the DAC resolution of a unitary converter system.

\begin{figure}[ht]
\begin{minipage}[b]{0.95\linewidth}
    \centering
    \includegraphics[width=8cm]{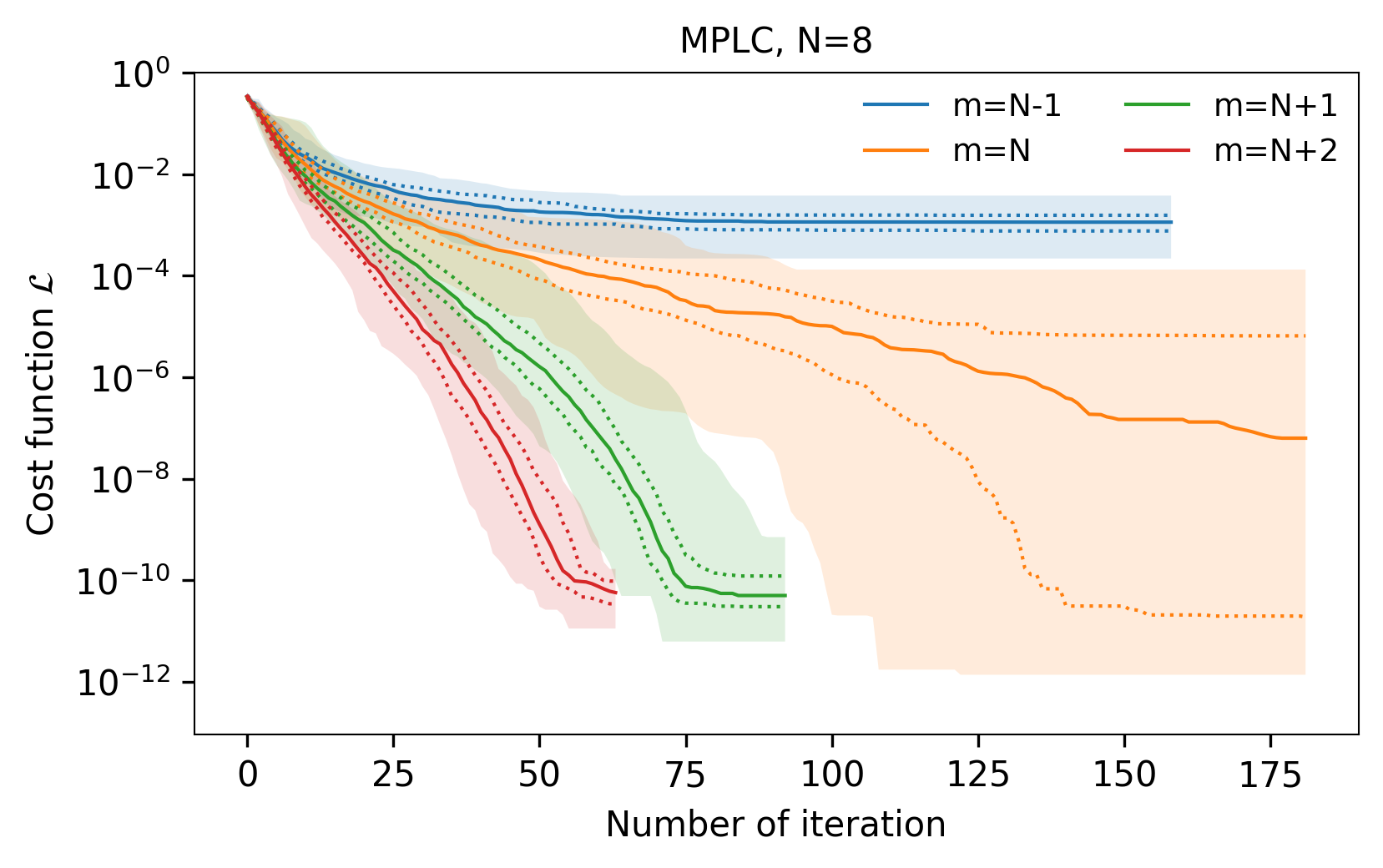}
\end{minipage}
  \caption{Convergence plots for the MPLC architecture using phase-insensitive distance with $N=8$. The 64 optimization trials are shown in the same manner as in Fig. \ref{fig:LBFGS_comp}.}
\label{fig:MPLC_phase_insensitive}
\end{figure}

\begin{figure}[ht]
\begin{minipage}[b]{0.95\linewidth}
    \centering
    \includegraphics[width=8cm]{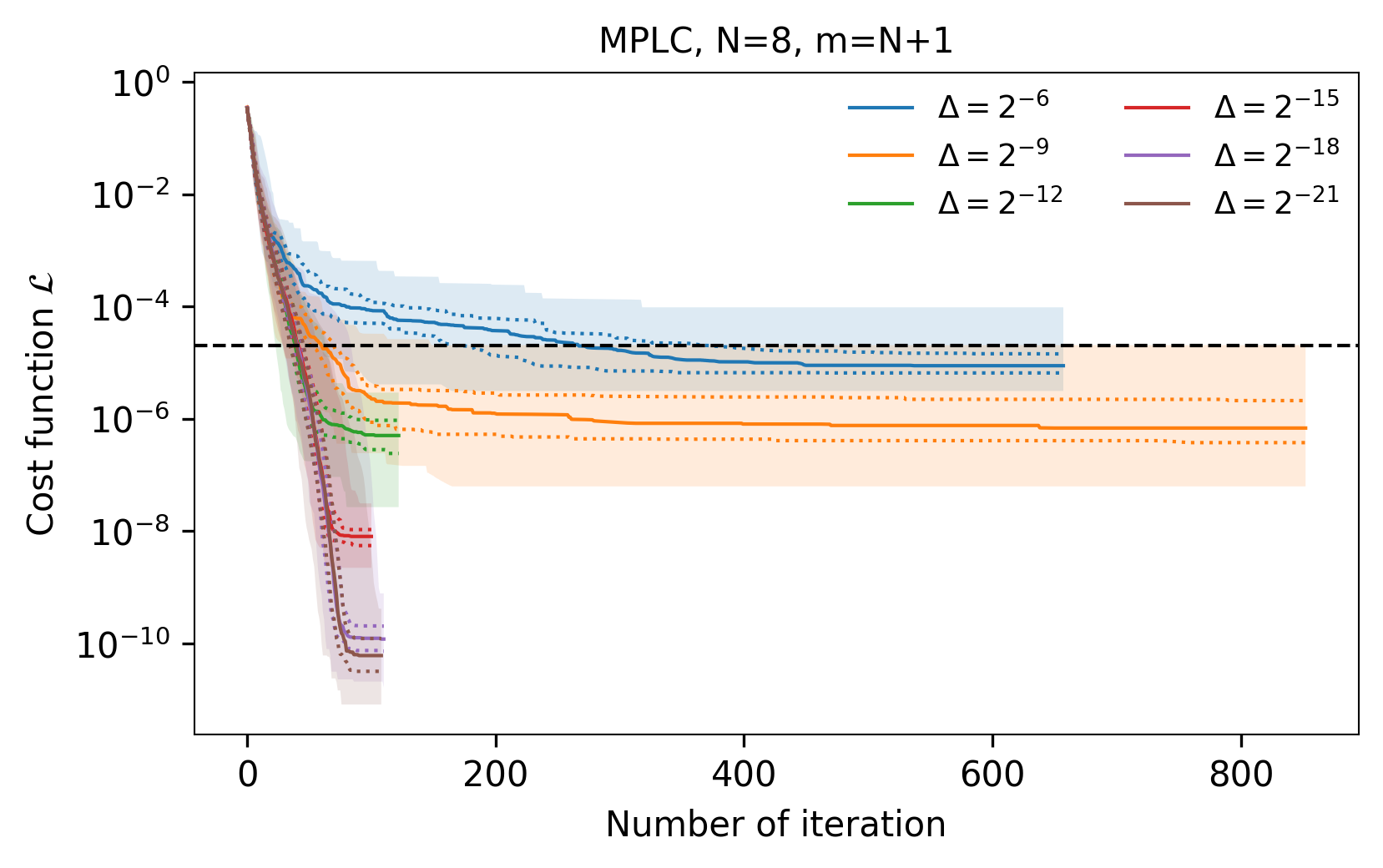}
\end{minipage}
  \caption{Comparison of final error when changing the accuracy of gradient approximation for the MPLC architecture using phase-insenstive distance. The black dashed line represents the optimization result in a previous study \cite{Tanomura2022PRA}, where $\mathcal{L}=2.0\times10^{-5}$ was achieved.}
\label{fig:approx_grad_acc_comp_MPLC_phase_insensitive}
\end{figure}

We studied the optimization property using the phase-insensitive distance. Fig. \ref{fig:MPLC_phase_insensitive} shows the optimization result obtained with an analytical gradient. The convergence plot is similar to the one shown in Fig. \ref{fig:LBFGS_comp}, in spite of the reduced degree of freedom. However, when using gradient approximation and comparing the accuracy dependence, the phase-insensitive distance shows a different result from the standard Frobenius-norm-based distance, as shown in Fig. \ref{fig:approx_grad_acc_comp_MPLC_phase_insensitive}. All the optimization results have a large variance, as opposed to the cases where $\Delta = 2^{-6}, 2^{-9}, 2^{-12}$ in Fig. \ref{fig:approx_grad_acc_comp_MPLC}. The finite difference $\Delta$ must be smaller than $2^{-18}$ to achieve an optimization result comparable to the one obtained using an analytical gradient.
The black dashed line shows the optimization result by simulated annealing in a previous study \cite{Tanomura2022PRA}. They achieved an error of $f_\mathrm{MSE}=\SI{-50}{\decibel}$, which corresponds to $\mathcal{L} = 2\times 10^{-5}$. This error can be achieved using our method with a finite difference of $\Delta \leq 2^{-9}$, and further improvement by orders of magnitude is possible with more accurate gradient approximation. When the accuracy is sufficient, our optimization method converges after about 100 iterations. To compare the speed of our method with a previous experimental report of MPLC architecture optimization using intensity detection \cite{Tanomura2020InP}, we also conducted optimization for a case with $N=4$ and $m=N+1$. The optimization converged after about 45 iterations with $\Delta = 2^{-18}$. As each iteration requires $4 \times (4+2)= 24$ evaluations, the total number of iterations required for convergence is 1080, representing a 23-fold speedup compared to the previous report ($\sim 25200$ evaluations).

\begin{figure}[ht]
\begin{minipage}[b]{0.95\linewidth}
    \centering
    \includegraphics[width=8cm]{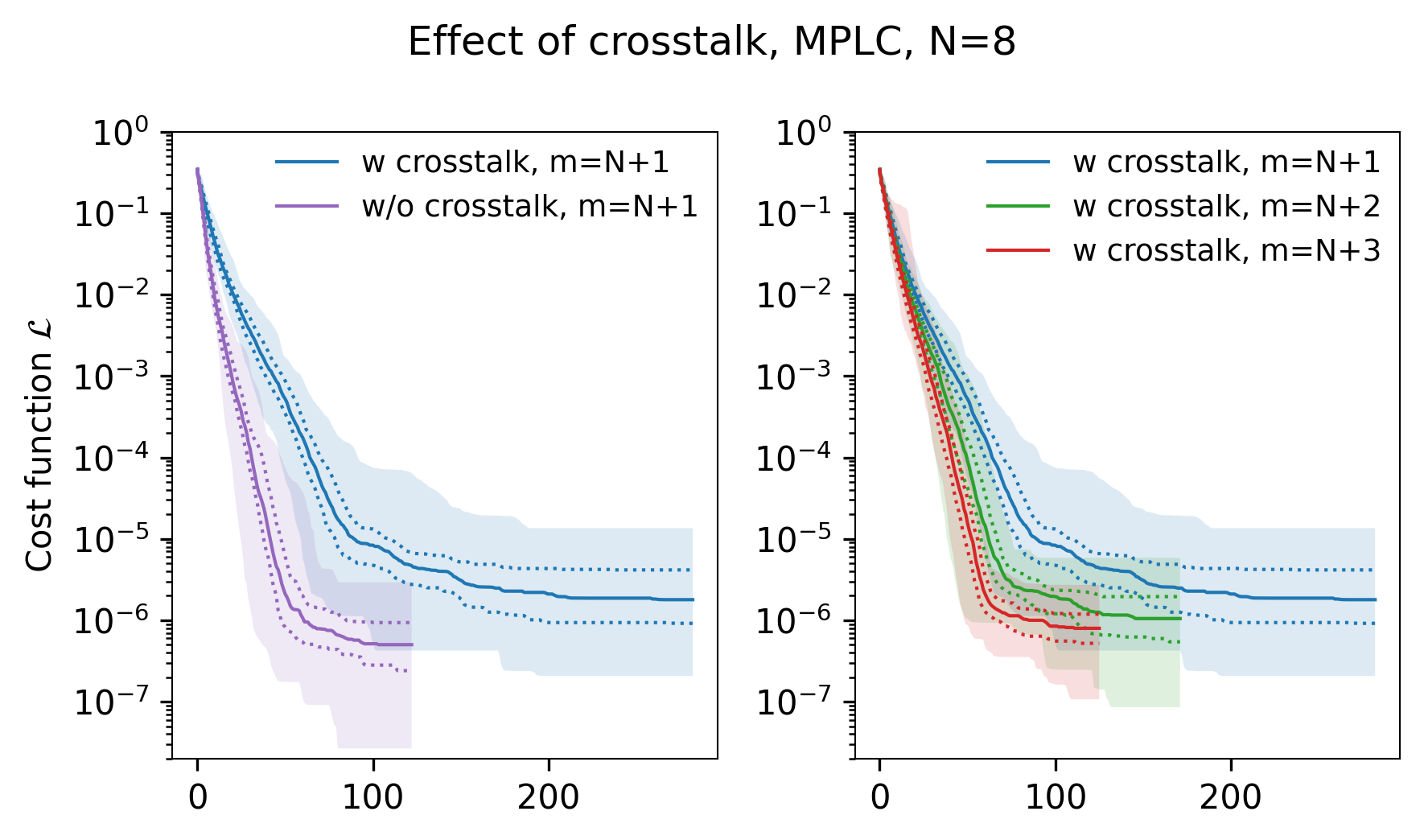}
\end{minipage}
  \caption{Comparison of final error under crosstalk using an approximated gradient with $\Delta=2^{-12}$ and phase-insenstive distance with $N=8$.}
\label{fig:approx_grad_acc_comp_crosstalk}
\end{figure}

The effect of crosstalk when using the phase-insensitive distance and the approximated gradient is shown in Fig. \ref{fig:approx_grad_acc_comp_crosstalk}. The gradient approximation was calculated using $\Delta=2^{-12}$. Although crosstalk caused a larger error and increased the number of iterations until convergence, the performance degradation can be mitigated by adding a few layers of additional redundancy. This result suggests that it may be possible to optimize the device end-to-end, including both matrix optimization and phase shifter calibration.

\section{Conclusion}
We proposed a fast and iteratively configurable MPLC architecture for realizing precise and fabrication-error-tolerant unitary transformation. Our numerical results show that adding a few redundant layers to the MPLC architecture significantly improves optimization behavior. We also examined the effect of artifacts, such as crosstalk, and found that the proposed architecture can be optimized end-to-end. In addition to proposing a new architecture and optimization method, we analyzed the distance between unitary matrices using the Frobenius norm. We introduced the concept of unimodality for functions on the unitary group and proved that the matrix distance using the Frobenius norm has this property. We also calculated the expected value and range of the matrix distance. We also introduced the phase-insensitive norm, which is useful for applications that only use intensity detections. We believe that this approach will enable the scalable and robust implementation of optical unitary converters and expand the use of photonic integrated circuits in various fields.

\begin{acknowledgments}
We wish to acknowledge Sho Yasui for the fruitful discussion. This work is supported by JST CREST Grant Number JPMJCR1872, Japan.
\end{acknowledgments}

\appendix
\section{Unimodality of the Frobenius norm}
\label{sec:proof_unimodality}
Here, we present an algebraic proof of the unimodality of the Frobenius norm. Let $X, U \in U(N)$.

\textbf{Theorem} \textit{Given U, if $f_U(X)=\norm{X-U}_F$ has a local minimum at some X, then it is the global minimum.}

\textbf{Proof.} The squared Frobenius norm can be expressed as $\norm{X-U}_F^2 = 2N - 2\Re\qty[\Tr\qty[U^\dag X]]$, and since this is a local minimum, the term $\Re\qty[\Tr\qty[U^\dag X]]$ is a local maximum. Consider $\Re\qty[\Tr\qty[U^\dag X]]$ as a function on the manifold $U(N)$.
We can investigate its critical points by examining the directional derivative of the function with respect to the tangent vector $X'$ at $X$.
The tangent vector $X'$ can be represented as $X'=ZX$, where $Z$ is a skew-Hermitian matrix and $X$ is any matrix on the unitary group $U(N)$. This is because the Lie algebra $\mathfrak{u}(N)$ of the unitary group $U(N)$ is composed of skew-Hermitian matrices
\footnote{Another proof for $X'=ZX$. Consider an identity $XX^\dag=I$. Taking the derivative of both sides, we get $X'X^\dag + X(X^\dag)' = O$. Then we can rewrite it as $X'X^\dag = -X(X^\dag)'=-(X'X^\dag)^\dag$. Let $Z = X'X^\dag$. Then, we have $Z=-Z^\dag$ which indicates $Z$ is a skew-Hermitian matrix. Since $Z=X'X^\dag$, we conclude that $X'=ZX$. \hspace{\fill}$\square$}.
When $X$ is at the critical point, then $\Re\qty[\Tr\qty[U^\dag X']] = 0$ is satisfied for any $X'$. Substituting $X'=ZX$, we obtain
\begin{equation}
\label{eq:appendix:ZXU}
\Re\qty[\Tr[U^\dag X']] = \Re\qty[\Tr[Z X U^\dag]] = 0.
\end{equation}
To further expand this equation, we consider two sets of special matrices $Z^1$ and $Z^2$, whose matrix $Z^1_{ij} \in Z^1, Z^2_{ij} \in Z^2$ is indexed by $1 \leq i, j \leq N$ with $i \neq j$. The $(k,l)$-th element of these matrices $\qty[\cdot]_{kl}$ is defined as follows:
\begin{align}
  [Z^1_{ij}]_{kl} &= \delta_{ik}\delta_{jl} - \delta_{il}\delta_{jk} \\
  [Z^2_{ij}]_{kl} &= i(\delta_{ik}\delta_{jl} + \delta_{il}\delta_{jk}) .
\end{align}
For example, each matrix set includes the following matrices:
\begin{align}
\newcommand{\myvdots}{\vphantom{\int^0}\smash[t]{\vdots}}
  Z^1_{12} = \begin{bmatrix}
    0 & 1 & 0 & \cdots & 0 \\
    -1 & 0 & & & \\
    0 &  &  & & \\
    \myvdots & & & & \myvdots \\
    0 & & & \cdots & 0
  \end{bmatrix},
  Z^2_{23} = \begin{bmatrix}
  0 & 0 & 0 & 0 & \cdots & 0 \\
  0 & 0 & i & 0 &  &  \\
  0 & i & 0 &   &  & \\
  0 & 0 &   &   &  & \\
  \myvdots & & & & & \myvdots \\
  0 &   &  & & \cdots & 0
  \end{bmatrix}.
\end{align}
After substituting $Z^1_{ij}, Z^2_{ij}, 1 \leq i, j \leq N$ for $Z$ in Eq. \ref{eq:appendix:ZXU}, we obtain
\begin{equation}
\left\{ \,
  \begin{aligned}
    & \Re\qty[\qty[X U^\dag]_{ij}] - \Re\qty[\qty[X U^\dag]_{ji}] = 0 \\
    & \Im\qty[\qty[X U^\dag]_{ij}] + \Im\qty[\qty[X U^\dag]_{ji}] = 0
  \end{aligned}
\right. ,
\end{equation}
which leads to $XU^\dag = (XU^\dag)^\dag$. Therefore,
\begin{equation}
\label{eq:appendix:XUXUI}
(XU^\dag)^2 = I.
\end{equation}
The unitary matrix $XU^\dag$ can be diagonalized using a regular matrix $V$, and a diagonal matrix $D$ whose diagonal elements $d_i \in \mathbb{C}$ satisfies $\abs{d_i}=1$ because $XU^\dag$ is a unitary matrix. We can express this diagonalization as $XU^\dag = VDV^{-1}$. Substituting $XU^\dag$ in Eq. \ref{eq:appendix:XUXUI}, we obtain $D = D^\dag$. Since $\abs{d_i}=1$, we have $d_i = \pm 1$. Now Consider the original local maximum term $\Re\qty[\Tr\qty[U^\dag X]]$. We can rewrite it as $\Re\qty[\Tr\qty[U^\dag X]] = \Re\qty[\Tr\qty[VDV^{-1}]] = \Re\qty[\Tr\qty[D]] = \Re\qty[\sum_i d_i]$. This value is obviously maximized when $d_i = +1$ for all $i$. If $d_i = -1$ for some $i$, we can rotate this value to $d_i=+1$ along the unit circle $\abs{c}=1$ in the complex plane and still achieve the maximum value. Therefore, if $\Re\qty[\Tr\qty[U^\dag X]]$ is a local maximum, it is also a global maximum. As a result, we now conclude that if $\norm{X-U}_F \geq 0$ is a local minimum, then it must also be a global minimum. Q.E.D.

\providecommand{\noopsort}[1]{}\providecommand{\singleletter}[1]{#1}%

\end{document}